\documentclass[longauth]{aa} 

\usepackage[varg]{txfonts} 
\usepackage{graphicx}
\usepackage{url}
\usepackage{natbib}
\usepackage{rotating}
\usepackage{longtable}
\usepackage{soul}
\usepackage{color}
\usepackage{natbib}
\usepackage{rotating}
\usepackage{lineno} 
\usepackage{xspace}
\usepackage{amsmath}
\usepackage{lineno} 

\bibpunct{(}{)}{;}{a}{}{,} 

\begin{document}%

   \title{TANAMI monitoring of Centaurus~A: The
complex dynamics in the inner parsec of an extragalactic jet}
   \author{
     C.~M\"uller\inst{\ref{affil:remeis},\ref{affil:wuerzburg}}
     \and M.~Kadler\inst{\ref{affil:wuerzburg}}
     \and R.~Ojha\inst{\ref{affil:nasa_gfsc}, \ref{affil:umbc}, \ref{affil:cua}}
     \and M.~Perucho\inst{\ref{affil:valencia}}
     \and C.~Gro\ss berger\inst{\ref{affil:remeis},\ref{affil:wuerzburg}}
     \and E.~Ros\inst{\ref{affil:mpifr},\ref{affil:valencia},\ref{affil:valencia1}}
     \and J.~Wilms\inst{\ref{affil:remeis}}
     \and J.~Blanchard\inst{\ref{affil:chile}}
     \and M.~B\"ock\inst{\ref{affil:mpifr}}
     \and B.~Carpenter\inst{\ref{affil:cua}}
     \and M.~Dutka\inst{\ref{affil:cua}}
     \and P.~G.~Edwards\inst{\ref{affil:csiro}}
     \and H.~Hase\inst{\ref{affil:bkg}}
     \and S.~Horiuchi\inst{\ref{affil:csiro_canberra}}
     \and A.~Kreikenbohm\inst{\ref{affil:remeis},\ref{affil:wuerzburg}}
     \and J.~E.~J.~Lovell\inst{\ref{affil:tasmania}}
     \and A.~Markowitz\inst{\ref{affil:ucsd},\ref{affil:remeis},\ref{affil:humboldt} }
     \and C.~Phillips\inst{\ref{affil:csiro}}
     \and C.~Pl\"otz\inst{\ref{affil:bkg}}
     \and T.~Pursimo\inst{\ref{affil:not}}
     \and J.~Quick\inst{\ref{affil:hartrao}}
     \and R.~Rothschild\inst{\ref{affil:ucsd}}
     \and R.~Schulz\inst{\ref{affil:remeis},\ref{affil:wuerzburg}}
     \and T.~Steinbring\inst{\ref{affil:wuerzburg}}
     \and J.~Stevens\inst{\ref{affil:csiro}}
     \and J.~Tr\"ustedt\inst{\ref{affil:wuerzburg}}
     \and A.K.~Tzioumis\inst{\ref{affil:csiro}}
              }

   \institute{
      Dr.\ Remeis Sternwarte \& ECAP, Universit\"at Erlangen-N\"urnberg,
      Sternwartstrasse 7, 96049 Bamberg, Germany \label{affil:remeis}
      \\\email{cornelia.mueller@sternwarte.uni-erlangen.de}
      \and
      Institut f\"ur Theoretische Physik und Astrophysik, Universit\"at
      W\"urzburg, Am Hubland, 97074 W\"urzburg, Germany
      \label{affil:wuerzburg}
      \and
      NASA Goddard Space Flight Center, Greenbelt, MD 20771, USA  \label{affil:nasa_gfsc}
       \and
      CRESST/University of Maryland Baltimore County, Baltimore, MD 21250, USA \label{affil:umbc}
       \and
      Catholic University of America, Washington, DC 20064, USA \label{affil:cua}
      \and
      Dept.\ d'Astronomia i Astrof\'isica, Universitat de Val\`encia,
      46100 Burjassot, Val\`encia, Spain \label{affil:valencia}
      \and
      Max-Planck-Institut f\"ur Radioastronomie, Auf dem H\"ugel 69, 53121
      Bonn, Germany \label{affil:mpifr}
      \and
       Observatori Astron\`omic, Universitat de Val\`encia, Parc
       Cient\'ific, C.\ Catedr\'atico Jos\'e Beltr\'an 2, 46980
       Paterna, Val\`encia, Spain \label{affil:valencia1}
        \and
       Departamento de Astronom\`ia Universidad de Concepci\'on, Casilla 160 C, Concepci\'on, Chile
     \label{affil:chile}
  \and
  CSIRO Astronomy and Space Science, ATNF, PO Box 76 Epping, NSW 1710, Australia \label{affil:csiro}
  \and
  Bundesamt f\"ur Kartographie und Geod\"asie, 93444 Bad K\"otzting, Germany \label{affil:bkg}
  \and
  CSIRO Astronomy and Space Science, Canberra Deep Space
  Communications Complex, P.O. Box 1035, Tuggeranong, ACT 2901,
  Australia \label{affil:csiro_canberra}
\and 
  School of Mathematics \& Physics, University of Tasmania, Private
  Bag 37, Hobart, Tasmania 7001, Australia \label{affil:tasmania} 
\and
  University of California, San Diego, Center for Astrophysics and
  Space Sciences, 9500 Gilman Dr., La Jolla, CA 92093-0424,
  USA \label{affil:ucsd}
  \and Alexander von Humboldt Fellow \label{affil:humboldt}
  \and
  Nordic Optical Telescope Apartado 474, 38700 Santa Cruz de La Palma
  Santa Cruz de Tenerife, Spain \label{affil:not}
  \and
  Hartebeesthoek Radio Astronomy Observatory, Krugersdorp, South
  Africa \label{affil:hartrao}
            }

   \date{\today}

 
  \abstract
   {Centaurus A (Cen~A) is the closest radio-loud active galactic nucleus.  Very
Long Baseline Interferometry (VLBI) enables us to study the spectral
and kinematic behavior of the radio jet--counterjet system on
milliarcsecond scales, providing essential information for jet
emission and propagation models.} 
{ In the framework of the TANAMI
monitoring, we investigate the kinematics and complex structure of
Cen~A on subparsec scales. We have been studying the evolution of the central parsec jet structure of Cen\,A
for over 3.5\,years. The proper motion analysis of individual jet
components allows us to constrain jet formation and propagation and to
test the proposed correlation of increased high-energy flux with jet ejection
events. Cen~A is an exceptional laboratory for such a detailed study because
its proximity translates to unrivaled linear resolution, where one
milliarcsecond corresponds to 0.018\,pc.}
   {As a target of the southern-hemisphere VLBI monitoring program
TANAMI, observations of Cen\,A are done approximately every six
months at 8.4\,GHz with the Australian Long Baseline Array (LBA) and
associated telescopes in Antarctica, Chile, New Zealand, and South
Africa, complemented by quasi-simultaneous 22.3\,GHz observations.}
   {The first seven epochs of high-resolution TANAMI VLBI observations
at 8.4\,GHz of Cen\,A are presented, resolving the jet on
(sub-)milliarcsecond scales. They show a differential motion of the
subparsec scale jet with significantly higher component speeds
farther downstream where the jet becomes optically thin. We determined
apparent component speeds within a range of $0.1\,c$ to $0.3\,c$ and identified long-term stable features. In combination with the
jet-to-counterjet ratio, we can constrain the angle to the line
  of sight to~$\theta\sim 12^\circ-45^\circ$.}
   {The high-resolution kinematics are best explained by a
spine-sheath structure supported by the downstream acceleration
occurring where the jet becomes optically thin. On top of the
underlying, continuous flow, TANAMI observations clearly resolve
individual jet features. The flow appears to be interrupted by an obstacle
causing a local decrease in surface brightness and circumfluent jet
behavior. We propose a jet-star interaction scenario to explain this
appearance.  The comparison of jet ejection times to high X-ray flux
phases yields a partial overlap of the onset of the X-ray emission and
increasing jet activity, but the limited data do not support a robust correlation. }

   \keywords{galaxies: active -- galaxies: individual (Centaurus A, NGC 5128) -- galaxies: jets -- techniques: high angular resolution}

  \authorrunning{C. M\"uller et al.}
  \titlerunning{TANAMI monitoring of Centaurus~A: The
complex dynamics in the inner parsec of an extragalactic jet}
   
   \maketitle
%
\section{Introduction}
Sample studies of jets of active galactic nuclei (AGN) reveal
that the overall jet flow follows a pre-existing channel, and
individual components can move at different speeds
\citep[e.g.,][]{Kellermann2004, Lister2009c}. The kinematics of
particular jet features can thus be described with a characteristic
speed primarily in the direction of the established jet, although
bends and twists can cause changes in the apparent motion vectors or
surface brightness due to Doppler boosting.  Higher speeds at higher
observing frequencies are found
\citep[e.g.,][]{Jorstad2001b,Jorstad2001a,Kellermann2004}, suggesting
that observations at different frequencies sample different parts of
the jet. These results can be explained by a spine-sheath-like
structure with a faster inner jet
\citep[e.g.,][]{Laing1999,Perucho2007a,Ghisellini2005,Cohen2007,
Tavecchio2008}.

Significant acceleration of components has been measured in a number
of AGN jets \citep{Homan2009a, Lister2013} that are both parallel and
perpendicular relative to the jet, although it is hard to distinguish
whether parallel acceleration is due to a change in the Lorentz factor
$\Gamma$ or a change in the angle to the line of sight. Overall, a
positive acceleration trend is found closer to the jet base at $\leq
15$\,pc, while deceleration occurs farther out.
Recent studies by \citet{Piner2012a} and \citet{Lister2013} have shown
that non-ballistic behavior, i.e., non-radial motion and acceleration,
are very common in blazar jets. These authors find a general trend toward
increasing speed down the jet for BL\,Lac objects and radio galaxies
and find that orientation effects cannot fully account for these speed
changes.

More detailed observations of extragalactic jets at the highest
possible resolution are required to study these effects. \object{Centaurus~A}
(Cen~A, PKS\,1322$-$428, NGC~5128) presents an ideal target for
investigating the innermost regions of AGN to substantiate these
proposed explanations.  At a distance of only 3.8\,Mpc
\citep{Harris2010}, the elliptical galaxy Cen~A hosts the closest
AGN. It exhibits powerful jets, which are observed in the radio and
X-ray regimes from subparsec scales up to hundreds of parsecs
\citep{Mueller2011a, Feain2011, Hardcastle2003, Clarke1992}.  The
morphology of these jets is consistent with Cen~A being a Fanaroff-Riley Type~I
radio galaxy \citep{Fanaroff1974}. Thanks to Cen~A's proximity, the
properties of its radio jet can be studied in unprecedented detail
on milliarcsecond (mas) scales using Very Long Baseline Interferometry
(VLBI), since at this distance an angular size of 1\,mas
translates into a linear scale of only $\sim$0.018\,pc. The formation
and propagation of extragalactic jets is still not completely
understood and requires high-resolution information to test and
compare them with theoretical models and simulations
\citep[e.g.,][]{Blandford1977,Blandford1982,Vlahakis2004,McKinney2005,McKinney2009,Tchekhovskoy2011,Fendt2014}.

The time evolution of the mas-scale jet of Cen~A has been studied well
for over 12\,years (1988 to 2000) with VLBI techniques by
\citet{Tingay1998b,Tingay2001b}. The jet kinematic has been measured
by tracking two prominent components with proper motion $\mu\sim
2\,\mathrm{mas}\,\mathrm{yr}^{-1}$, corresponding to an apparent
velocity $\beta_\mathrm{app}\sim 0.12$ (where $\beta=v/c$) for the
inner mas-scale jet (up to $\sim 30$\,mas distance from the
core). These observations reveal a stationary component at
$\sim$4\,mas distance from the core, with a flux density of $\sim
0.5\pm0.3$\,Jy. \citet{Horiuchi2006} report on a single-epoch
space-VLBI observation of Cen~A at 5\,GHz showing a well-collimated
jet, with an intriguingly similar structure to our previously
presented results \citep[][hereafter Paper~I]{Mueller2011a}, which
suggests generally stable conditions for the formation and propagation
of the jet.

On larger scales, up to 100\,pc from the core, \citet{Hardcastle2003}
measured a bulk velocity of $\beta_\mathrm{app}\sim 0.5$ using Very
Large Array (VLA) and \textsl{Chandra} observations. The presence of
radio and X-ray emitting knots could be explained by the interaction
of the jet with stars or gas clouds in the host galaxy
\citep{Hardcastle2003,Tingay2009}. 
\citet{Worrall2008a} observe a steeper spectrum
for components in the outer layers of the jet, which suggests a
spine-sheath character for the 100-pc-scale jet.

Hard X-ray observations reveal strong absorption below
\mbox{$2-3$\,keV}, above which the spectrum in the hard X-rays can be
modeled with a power law ($\Gamma\sim 1.8$) and narrow fluorescence
lines \citep{Evans2004a,Markowitz2007}.  Furthermore, observations by
\citet{Fukazawa2011} indicate jet emission in this energy band.
\citet{Tingay1998b} discussed a possible correlation of the increase
in X-ray flux with the jet component ejection, although the limited
data precluded firm conclusions.

In the framework of the VLBI monitoring program TANAMI
\citep[\textsl{Tracking Active Galactic Nuclei with Milliarcsecond
Interferometry},][]{Ojha2010a}, Cen~A has been monitored at 8.4\,GHz
and 22.3\,GHz approximately twice a year since 2007. We previously
presented the first dual-frequency TANAMI observation of this source, which
resulted in a very detailed jet image at highest possible linear
resolution (Paper~I). These images reveal a highly collimated jet at a distance of a
few light days from the black hole. 
The broadband spectral energy distribution (SED) can be
described with a single-zone
synchrotron/synchrotron self-Compton model \citep{Abdo2010_cenacore},
though the spectral index distribution of
the subparsec scale jet (Paper~I) indicates multiple possible emission sites of
gamma rays.

A detailed scrutiny of possible high-energy emission
processes in Cen~A's inner parsec jet system is very important given
the suggested positional coincidence of Cen~A with ultra-high-energy
cosmic rays observed by the Pierre Auger Observatory
\citep[e.g.,][]{Romero1996,Honda2009,Anchordoqui2001,Anchordoqui2011,Kim2013}. 
Furthermore, it has recently been shown that the brightest
extragalactic jets inside the fields of the first two PeV neutrino
events detected by IceCube \citep{IceCube2013,eb} are capable, from a
calorimetric point of view, of producing the observed neutrino flux
\citep{Krauss2014}.  In this context, it is intriguing that Cen~A
positionally coincides with the recently reported third PeV neutrino
event \citep{bigbird}.

Here we report on the evolution of the mas-scale jet structure of
Centaurus~A at 8.4\,GHz. The paper is organized as follows. In
Sect.~\ref{sec:obs} we summarize the observations and data reduction
procedure. We then present the
high-resolution images at 8.4\,GHz in Sect.~\ref{sec:results}, concentrating on the time
evolution of the jet (Sect.~\ref{susec:evolution}). We discuss these
results in Sect.~\ref{sec:discussion} and conclude with an overall
picture in Sect.~\ref{sec:conclusion}.

\section{Observations and data reduction}\label{sec:obs}
The TANAMI monitoring program \citep[][]{Ojha2010a} has made seven
VLBI observations of Cen~A at 8.4\,GHz between 2007 November and 2011
April. TANAMI uses the Australian Long Baseline Array (LBA) and
additional radio telescopes in Antarctica, Chile, New Zealand, and
South Africa.  Data in the 8.4\,GHz band were recorded in single polarization
mode with two-bit sampling and correlated
on the DiFX software correlator \citep{Deller2007, Deller2011} at
Curtin University in Perth, Western Australia. They were calibrated
using standard procedures in AIPS \citep{Greisen2003} as described in
\citet{Ojha2010a}. A log of observations and the corresponding image
parameters can be found in Table~\ref{table:1}. This includes the
specific array configuration of each observation, which in general,
varies from epoch to epoch. Owing to the contribution of different
non-LBA telescopes at different epochs, the $(u,v)$-coverage, array
sensitivity, and angular resolution vary between the different
observations.  The imaging and self-calibration process of TANAMI data
is performed using the \texttt{CLEAN}-algorithm implemented in the
program \texttt{DIFMAP} \citep{Shepherd1997}. The major constraint on
the image fidelity at 8.4\,GHz is the lack of intermediate baselines
between the LBA and transoceanic antennas (Paper~I). Further details
of the TANAMI data reduction and imaging are reported in
\citet{Ojha2010a} and, specifically for Cen~A, in Paper~I.

\begin{table*}
\caption{Details of the 8.4\,GHz TANAMI observations of Centaurus~A}             
\label{table:1}      
\resizebox{\textwidth}{!}{
\begin{tabular}{c c c c c c c c r }    
\hline\hline 
Obs.~date & Array configuration\tablefootmark{a} & $S_\mathrm{peak}$\tablefootmark{b} & RMS\tablefootmark{b} & $S_\mathrm{total}$\tablefootmark{b} & $\mathrm{S_\mathrm{inner~jet}}$\tablefootmark{c} &$b_\mathrm{maj}$\tablefootmark{d} & $b_\mathrm{min}$\tablefootmark{d} & P.A.\tablefootmark{d} \\
(yyyy-mm-dd)  &                  & (Jy beam$^{-1}$) & (mJy
beam$^{-1}$) & (Jy) & (Jy) & (mas) & (mas) & ($^\circ$) \\
\hline
2007-11-10 & AT-MP-HO-HH-CD-PKS        & $0.60\pm0.09$        &
$0.40\pm0.06 $ & $2.6\pm0.4$ & $2.5\pm0.4$& 1.64 & 0.41 & 7.9 \\
2008-06-09 & AT-MP-HO-HH-CD-PKS        & $1.06\pm0.16$        &
$0.63\pm0.09$ & $3.1\pm0.5$ & $2.7\pm0.4$ &2.86 & 1.18 & $-12.7$\\
2008-11-27 & TC-OH-AT-MP-HO-CD-PKS-DSS43 & $0.74\pm0.11$        & $0.37\pm0.06 $ & $4.0\pm0.6 $ & $3.1\pm0.5$& 0.98 & 0.59 & 31.4 \\
2009-09-05 & TC-OH-AT-MP-HO-CD-PKS-DSS43 & $0.76\pm0.11$        & $0.45\pm0.07 $ & $4.0\pm0.6 $ & $3.0\pm0.4$&2.29 & 0.58 & 15.6 \\
2009-12-13 & TC-AT-MP-HO-CD-PKS          & $1.03\pm0.15$        & $0.18\pm0.03 $ & $3.8\pm0.6 $ & $3.0\pm0.4$&3.33 & 0.78 & 26.3 \\
2010-07-24 & TC-AT-MP-HO-CD-PKS          & $1.21\pm0.18$        &
$0.38\pm0.06 $ & $4.2\pm0.6$ & $3.3\pm0.5$& 2.60 & 0.87 & 21.4\\
2011-04-01 & TC-WW-AT-MP-HO-HH-CD-PKS-DSS43 & $0.63\pm0.09$   &
$0.31\pm0.05 $ & $5.1\pm0.8 $ & $3.8\pm0.6$& 2.31 & 0.51 & $-0.7$\\
\hline  
\end{tabular}
}
\tablefoot{
\tablefoottext{a}{AT: Australia Telescope Compact Array, CD: Ceduna,
HH: Hartebeesthoek, HO: Hobart, MP: Mopra, OH: GARS/O'Higgins, PKS:
Parkes, TC: TIGO, DSS43: NASA's Deep Space Network Tidbinbilla
(70\,m), WW: Warkworth}
\tablefoottext{b}{Peak flux density, RMS noise level and total flux density in the \texttt{CLEAN}-image.}
\tablefoottext{c}{TANAMI 8.4\,GHz flux density
of inner jet defined by $-2\,\mathrm{mas}\lesssim \mathrm{RA}_\mathrm{relative}\lesssim
15$\,mas and adopting an uncertainty of 15\% (see
Sect.~\ref{susec:LC}).}
\tablefoottext{d}{Major and minor axes and position angle of restoring
beam.}
}
\end{table*}

%
\section{Results}\label{sec:results}
\subsection{High-resolution imaging}\label{susec:imaging}

\begin{figure*}
\includegraphics[width=0.48\textwidth]{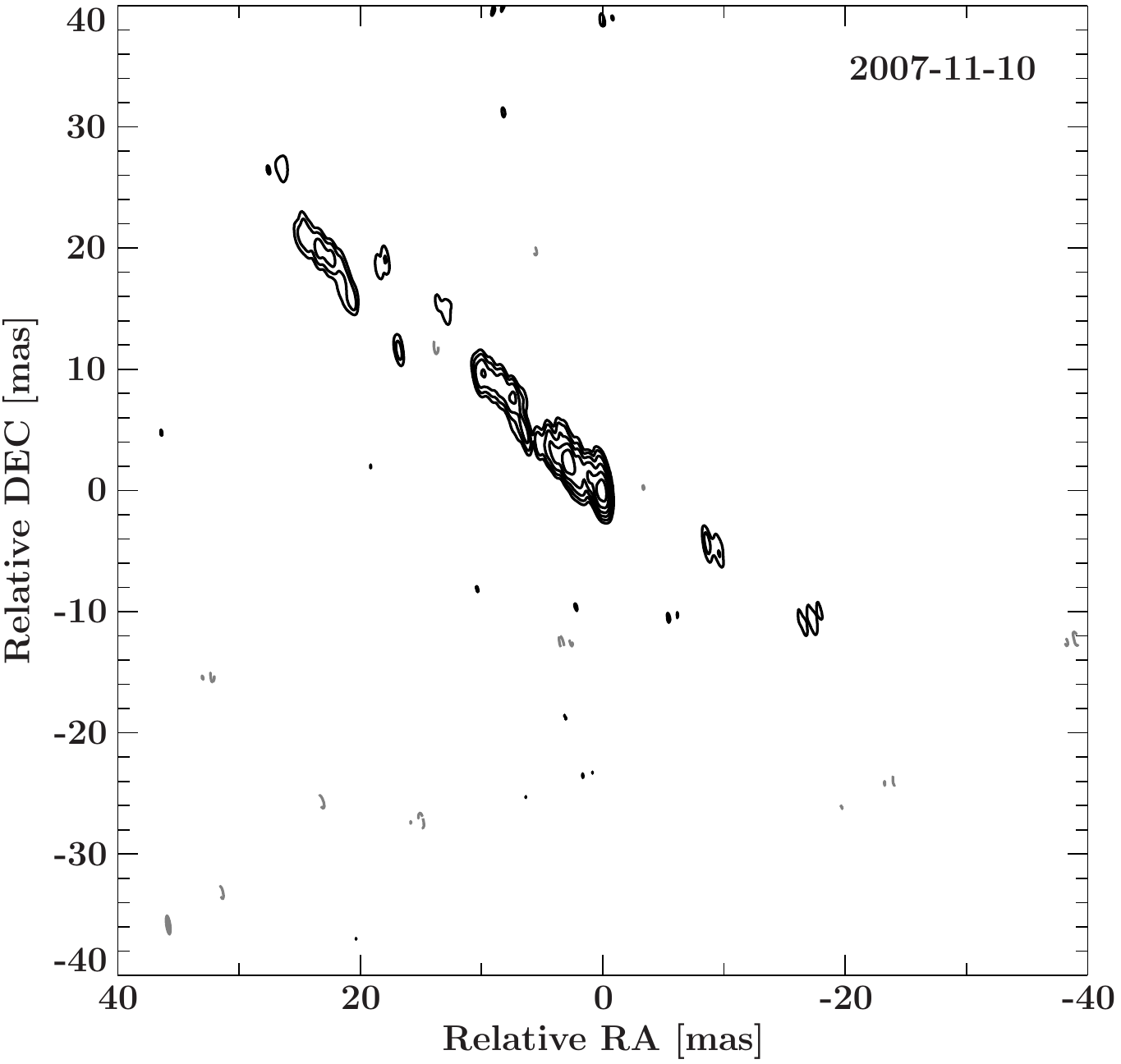}\hfill
\includegraphics[width=0.48\textwidth]{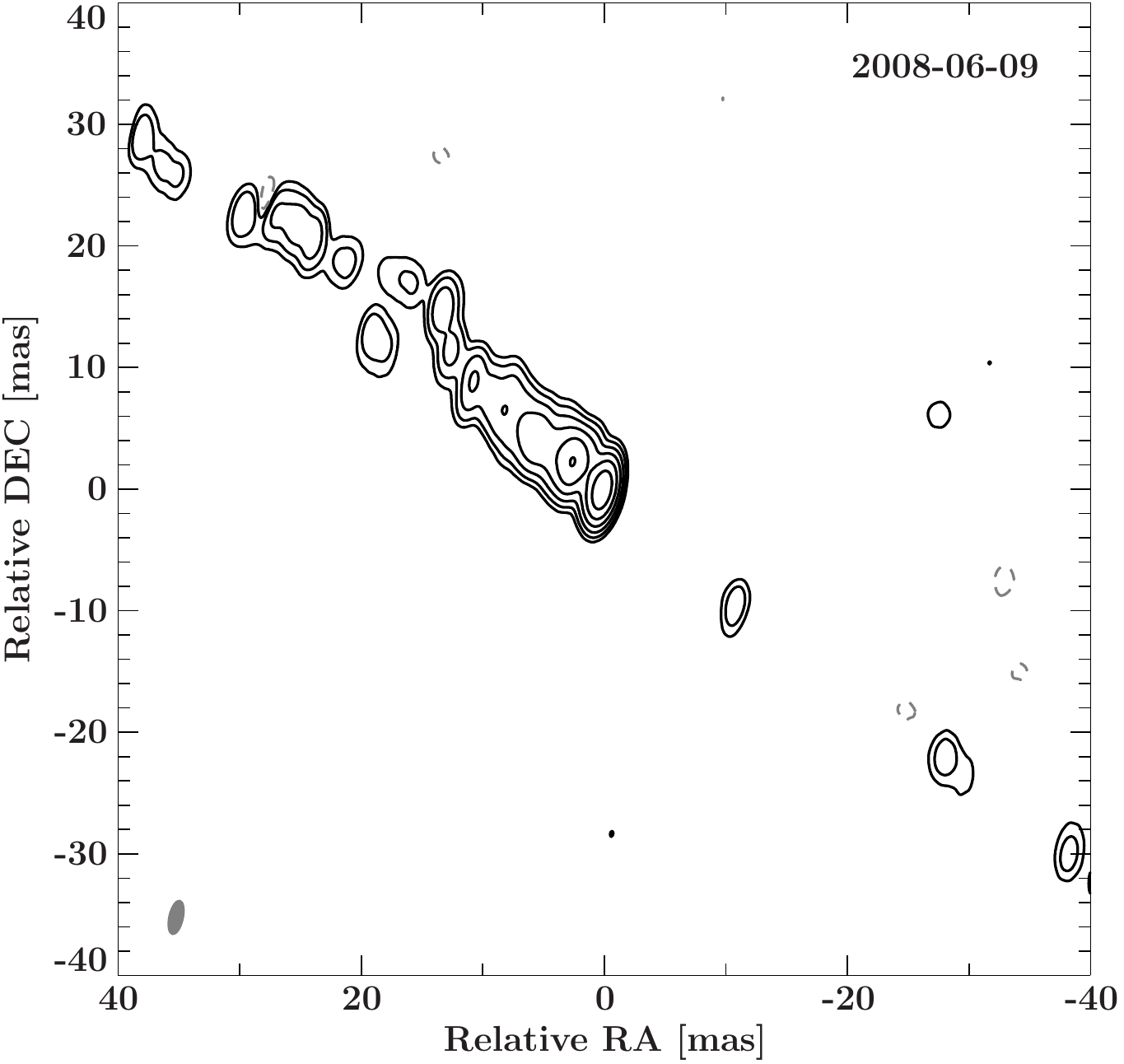}
\includegraphics[width=0.48\textwidth]{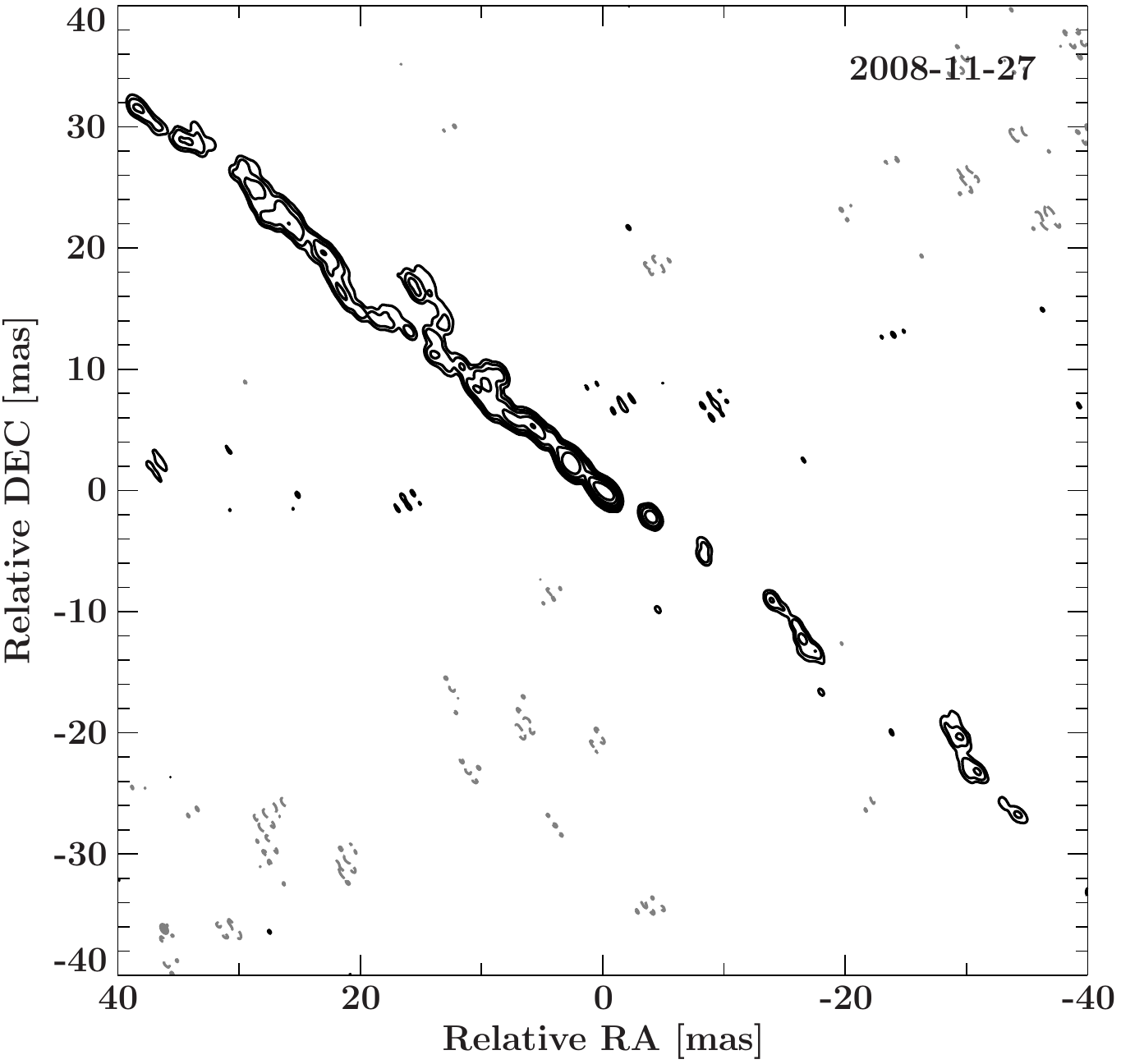}\hfill
\begin{minipage}[b]{0.48\textwidth}
  \caption{Contour images of the first three 8.4\,GHz TANAMI
observations with natural weighting. The black contours indicate the
flux density level, scaled logarithmically and increased by a factor
of 3, with the lowest level set to the $3\sigma$-noise-level (for more
details see Table~\ref{table:1}). Negative contours are shown in
gray. From top left to bottom right: 2007 November
\citep[][]{Ojha2010a}, 2008 June, 2008 November (Paper~I). The FWHM of
the restoring beams for each observation is shown as a gray ellipse in
the lower left corner.  1\,mas in the image corresponds to
$0.018$\,pc. \label{fig:images} }
\end{minipage}
 
\end{figure*}

\begin{figure*}
\includegraphics[width=0.48\textwidth]{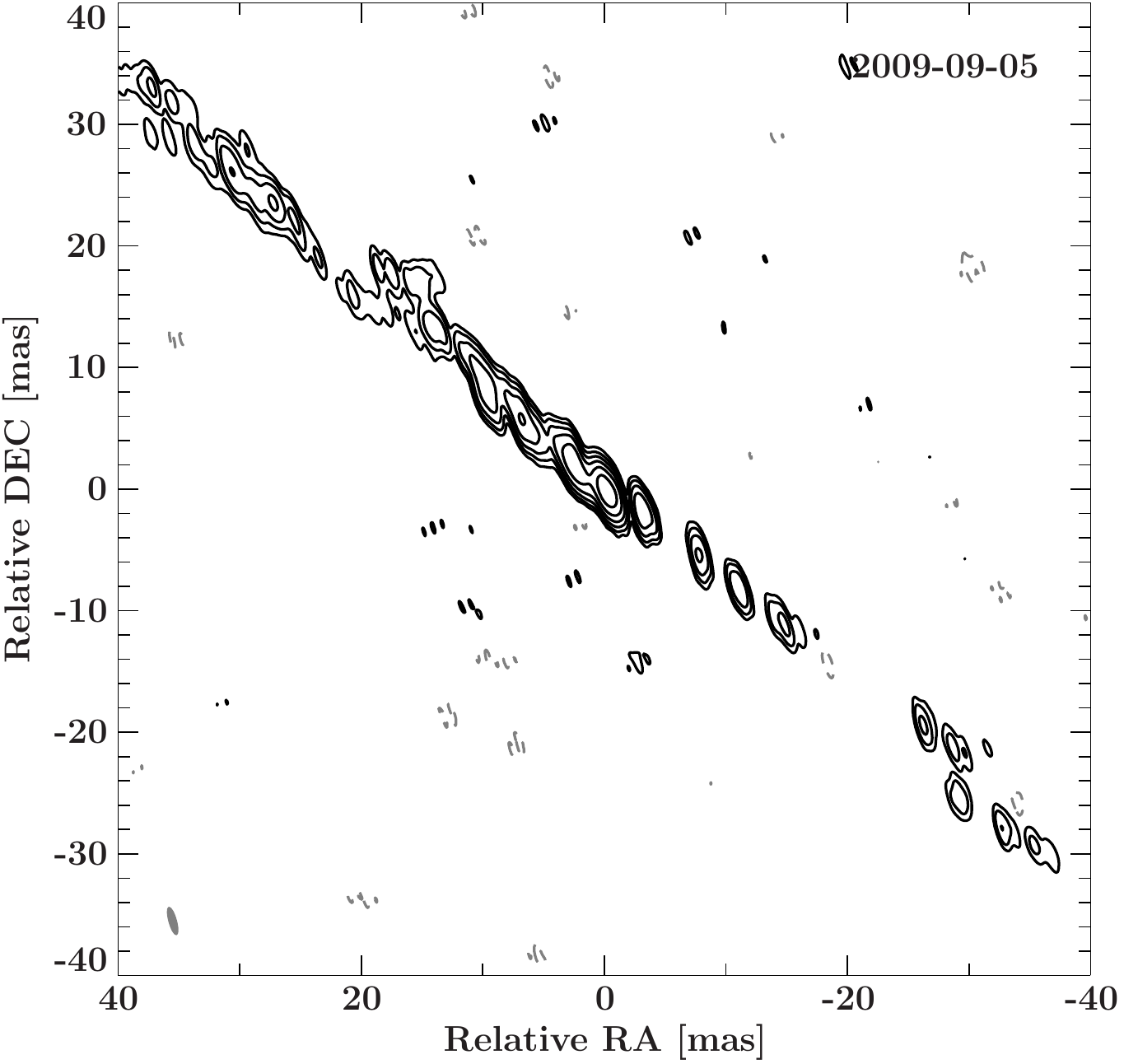}\hfill
\includegraphics[width=0.48\textwidth]{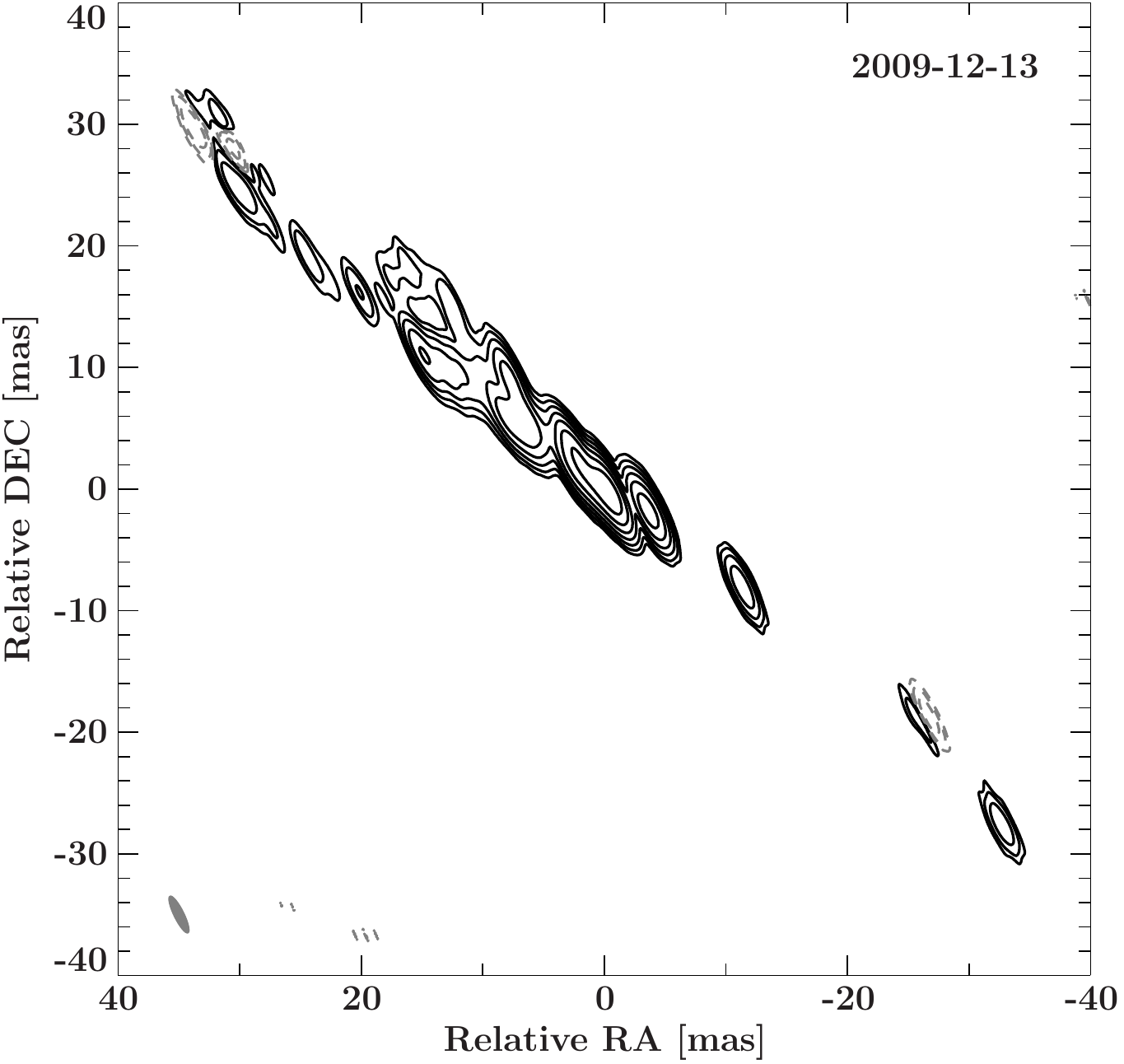}
\includegraphics[width=0.48\textwidth]{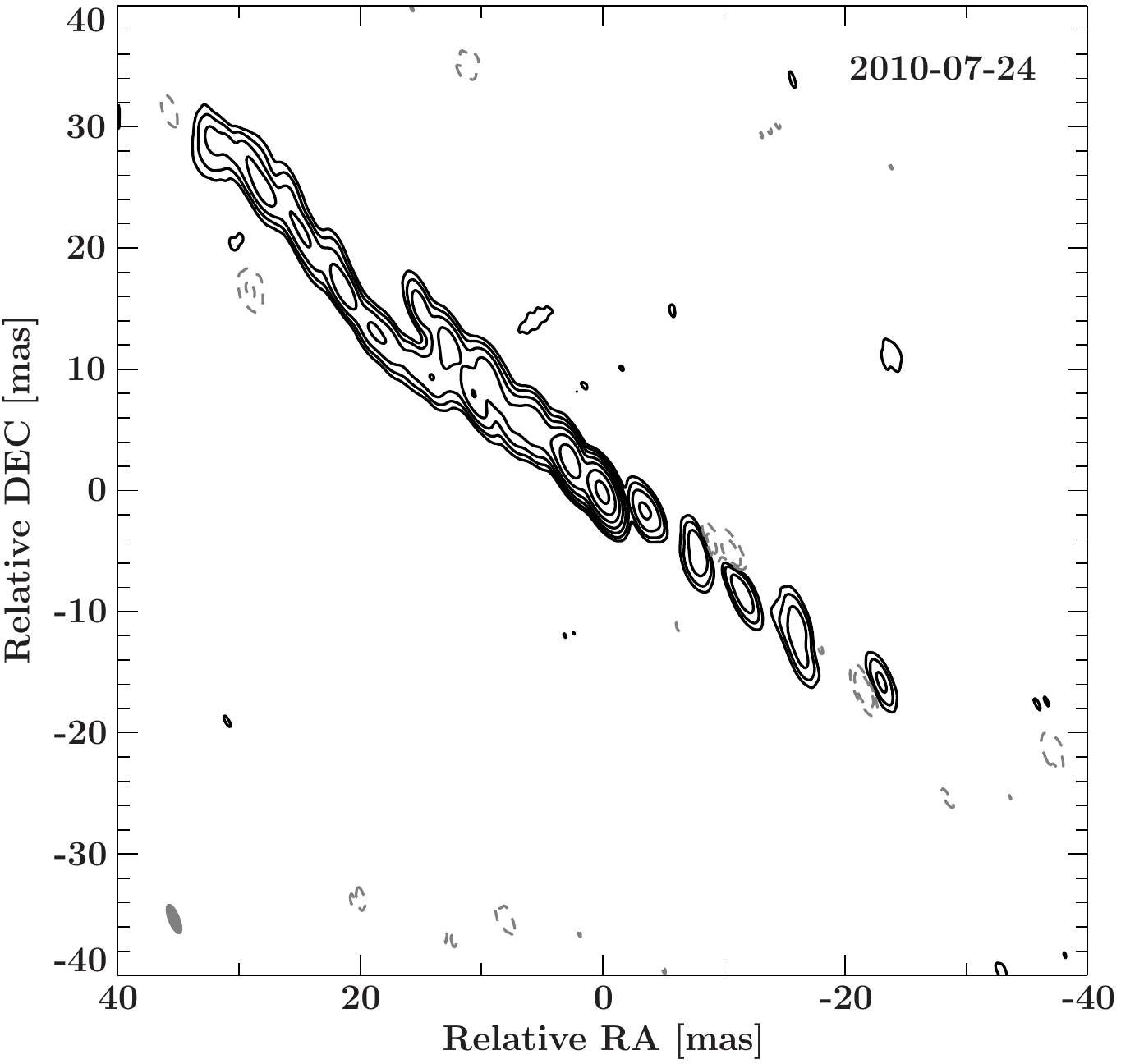}\hfill
\includegraphics[width=0.48\textwidth]{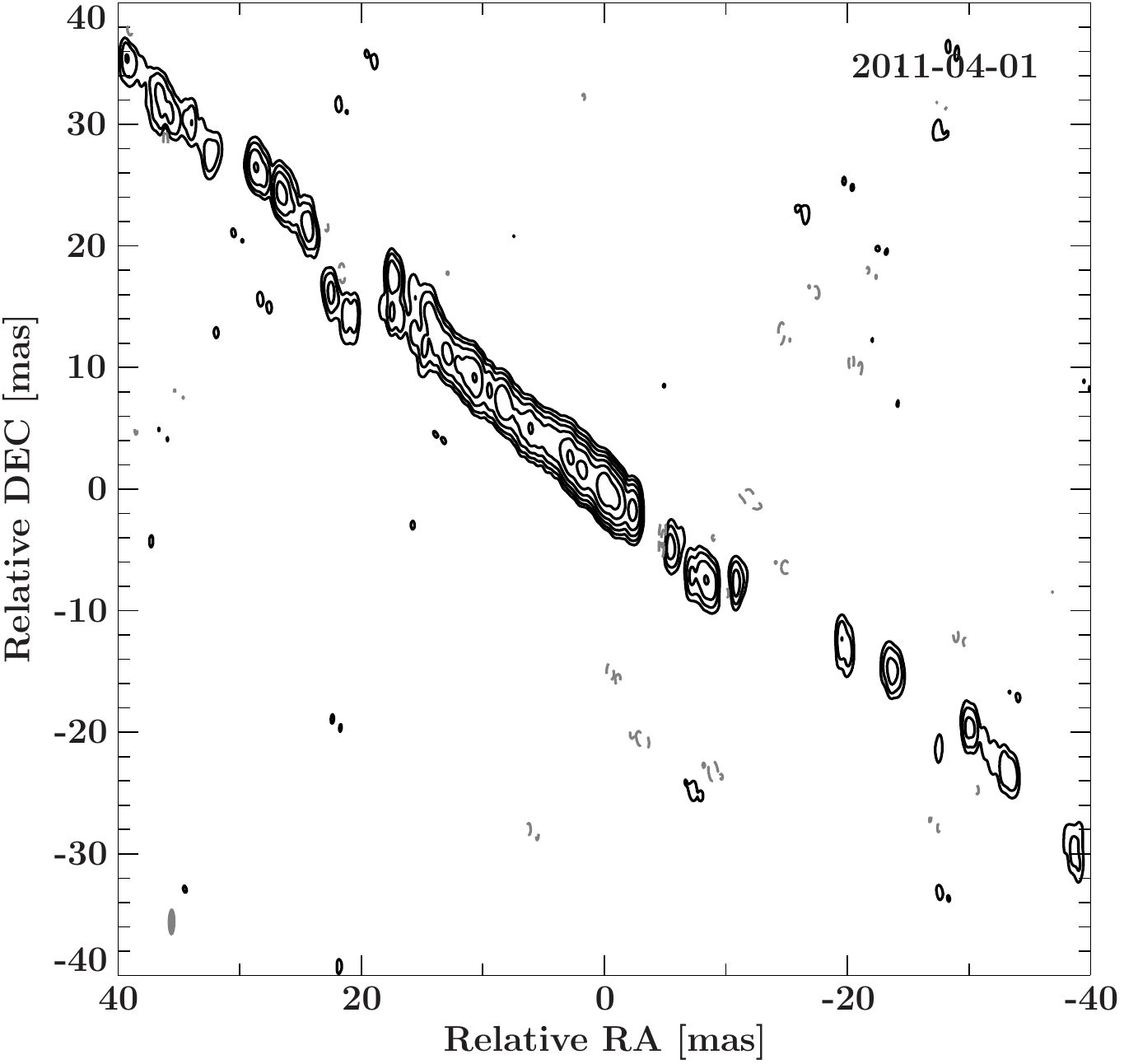}
\caption{Same as Fig.~\ref{fig:images} for the $4^\mathrm{th}$ to
  $7^\mathrm{th}$ TANAMI observations. From top left to bottom right:
      2009 September, 2009 December, 2010 July, and 2011 April. \label{fig:images2} } 
\end{figure*}

Figures~\ref{fig:images} and~\ref{fig:images2} show the naturally weighted images of the
first TANAMI observations at 8.4\,GHz with mas resolution\footnote{Figures~\ref{fig:images} and~\ref{fig:images2}
  are available in electronic form
at the CDS via anonymous ftp to cdsarc.u-strasbg.fr (130.79.128.5)
or via \url{http://cdsweb.u-strasbg.fr/cgi-bin/qcat?J/A+A/}.}. The image
parameters and observation characteristics are listed in
\mbox{Table~\ref{table:1}}. The comparison of the images with
different angular resolutions and dynamic ranges (defined as ratio of
peak brightness to three times the RMS noise level) from $\sim$500 to
$\sim$1900 allows us to further constrain the position and motion of
individual bright jet features.

These images display the jet of Centaurus~A in unprecedented
detail, clearly resolving particular jet features. The jet-counterjet
system is detected in all images and is already well collimated on
subparsec scales. We measured a jet opening angle of
$\theta\lesssim12^\circ$ on scales of $\lesssim 0.3$\,pc (Paper~I).

The jet emission is detected up to a maximum extent of $\sim70$\,mas
($\approx 1.3$\,pc) from the image phase center and shows an overall
straight, well-collimated morphology without any major bends.  The
counterjet is substantially weaker than the jet at a much lower
signal-to-noise ratio. It is as strongly collimated as the jet.

We determine a mean position angle of $\mathrm{P.A.}\sim 50^\circ$.
The stacked flux density profile along this jet axis
(Fig.~\ref{fig:fluxcuts}) reveals a declining surface brightness
distribution, with resolved individual emission humps.

Assuming an overall calibration uncertainty of 15\% (see
Paper~I for details), the flux density of the inner ($\sim$20\,mas)
jet shows an increasing trend over the period of observations reported
here ($\sim$3.5\,years). In particular, we measure a flux density
increase of $\Delta S_\mathrm{inner~jet}=1.3\pm0.7$\,Jy
between 2007 and 2011 (see Table~\ref{table:1},
Fig.~\ref{fig:LC} and discussion in Sect.~\ref{susec:LC}) in this
region.

We identify the VLBI core of the jet at 8.4\,GHz as the brightest
feature in the map. This result is based on the spectral index imaging
in Paper~I, as well as on the comparison of individual epochs, 
i.e., the remarkably similar overall jet structure. 
The brightest jet knot appears as a
pronounced, isolated feature in the stacked profile (see
Fig.~\ref{fig:fluxcuts}). We assume that it is stationary and 
considered it to be the 8.4\,GHz core.

The second brightest jet feature, an isolated component downstream
next to the core at a distance of $\sim$3.5\,mas (see
Fig.~\ref{fig:fluxcuts}), is found to be stationary with respect to
the core (see Sect.~\ref{susec:evolution} and~\ref{susec:jstat}). It
most probably corresponds to the stationary component C3 detected by
\citet{Tingay2001b}. Preceding this bright stationary feature (labeled
as $\mathrm{J_{stat}}$, see Table~\ref{table:2} and
Fig.~\ref{fig:timeevolution}), i.e., between this component and the core component, a
clear decrease in brightness is measured. Figure~\ref{fig:fluxcuts}
shows that this dip is also not moving over our monitoring period.
However, in epoch 2008.9, an additional component, later associated
with J10, is found in this region.

In Paper~I we reported on a possible widening and subsequent narrowing
of the jet appearing downstream at a distance of $\sim25$\,mas from
the core in our 2008 November image. This bifurcation or ``tuning
fork''-like emission structure around $25.5\pm2.0$\,mas is seen in all
other 8.4\,GHz TANAMI images, although partially masked by the varying
uv-coverage in some images.  The flux density profile of the stacked
image (Fig.~\ref{fig:fluxcuts}) illustrates the persistent local
flux-density minimum best. It manifests as a remarkably sharp dip in the
profile, while most other features are washed out by component motion
(see Sect.~\ref{susec:fork}).  This indicates that the ``tuning fork''
is a stationary feature in the central-parsec jet of Cen~A between
2007 and 2011. The overall jet flux density distribution is
declining with distance and drops where the jet widens. In
Sect.~\ref{susec:fork} we discuss the nature of this
particular jet structure further.
 
\subsection{Tapered data analysis}\label{susec:taper}
The high-resolution TANAMI observations of Cen~A are clearly resolving
the (sub-)parsec scale jet along and perpendicular to its axis. The
overall appearance suggests that we detect substructure in the
underlying jet flow.

To test this assumption and to connect our results to
previous VLBI studies of the Cen~A jet at 8.4\,GHz with a
  $\sim (3-15)$\,mas angular resolution \citep{Tingay1998b,
Tingay2001b}, we applied a $(u,v)$-taper to each dataset, so that the
angular resolution compares to previous measurements. As an example,
Fig.~\ref{fig:taper} shows the naturally weighted image at \mbox{(sub-)mas}
resolution, as well as the tapered and restored image with comparable
resolution to the earlier observations by \citet{Tingay2001b}. It is
obvious that the TANAMI array is resolving small jet substructures,
while the main emission zones, which appear as single jet knots in
these previous (lower resolution) images, look remarkably
similar. This result shows that the observed high-resolution
structures are consistent with earlier measurements. 

\citet{Tingay2001b} discuss the evolution of three individual jet
knots C1, C2, and C3. The outer components C1 and C2 have a mean
component motion of $\sim 2\,\mathrm{mas}\,\mathrm{yr}^{-1}$, while C3
appears to be stationary \citep{Tingay1998b, Tingay2001b}. Adopting
the respective reported apparent motions and uncertainties, we can
determine their expected positions at the time of our TANAMI
observations.  Figure~\ref{fig:taper} shows that these extrapolated
positions match well with prominent emission regions in the tapered
TANAMI image within the uncertainties given by the velocity
measurement by \citet{Tingay2001b}. We can therefore associate these
components with features detected in the TANAMI images (see
Sect.~\ref{susec:evolution}).  This comparison is further discussed in
Sect.~\ref{sec:discussion}.

\subsection{Time evolution of the mas-scale jet}\label{susec:evolution}
The high spatial resolution and the good sampling of the TANAMI
monitoring of Cen~A at 8.4\,GHz allow us to study the time evolution
of the closest extragalactic jet in unprecented detail. To
parametrize individual jet features and to track their evolution with
time, we fit Gaussian emission model components to the
self-calibrated visibility data (see Sect.~\ref{susec:imaging}) using
the \texttt{modelfit} task in \texttt{DIFMAP} \citep{Shepherd1997}.
The best-fit parameters for each TANAMI 8.4\,GHz observation of Cen~A
are listed in Table~\ref{table:2}.  The identified jet components,
which can be tracked over several epochs, are explicitly labeled.  The
models for every single epoch were constructed following the same
procedure. The self-calibrated visibility data were fit with
sufficient Gaussian model components to describe the prominent bright
jet features, leading to final models with $\chi^2\sim2232$
(d.o.f.=$2350$)\footnote{The number of degrees of freedom is
determined as the number of real and imaginary part of complex
visibilities minus the number of free model parameters.} to
$\sim23906$ (d.o.f$=6846$, see Table~\ref{table:2}).
Figure~\ref{fig:timeevolution} summarizes the result of the kinematic
study showing the contour images overlaid by the corresponding
Gaussian model components\footnote{Owing to the lower SNR of
the counterjet, no kinematic measurements could be obtained.}.  The
brightest upstream feature, identified as the core (see
Sect.~\ref{susec:imaging}), was taken as a reference position and all
positions are measured with respect to it.

We performed a statistical error calculation for the model fit
parameters by interfacing \texttt{DIFMAP} with the data-reduction
package \texttt{ISIS} \citep{Houck2000}. This approach allows us to
use the functionality contained in this general purpose system to
efficiently determine the parameter values and their uncertainties
based on the $\chi^2$ statistics. The error calculation was performed
for each parameter. We find that the systematic uncertainties dominate
(typically by more than an order of magnitude) over the statistical
errors even if cross-coupling between adjacent components is
considered. Therefore, we adopt the semi-major component axis as an
estimate of positional uncertainties of individual emission features
and conservatively estimated 15\,\% calibration uncertainties (see
Paper~I) for component flux densities.

\begin{figure}
\includegraphics[width=\hsize]{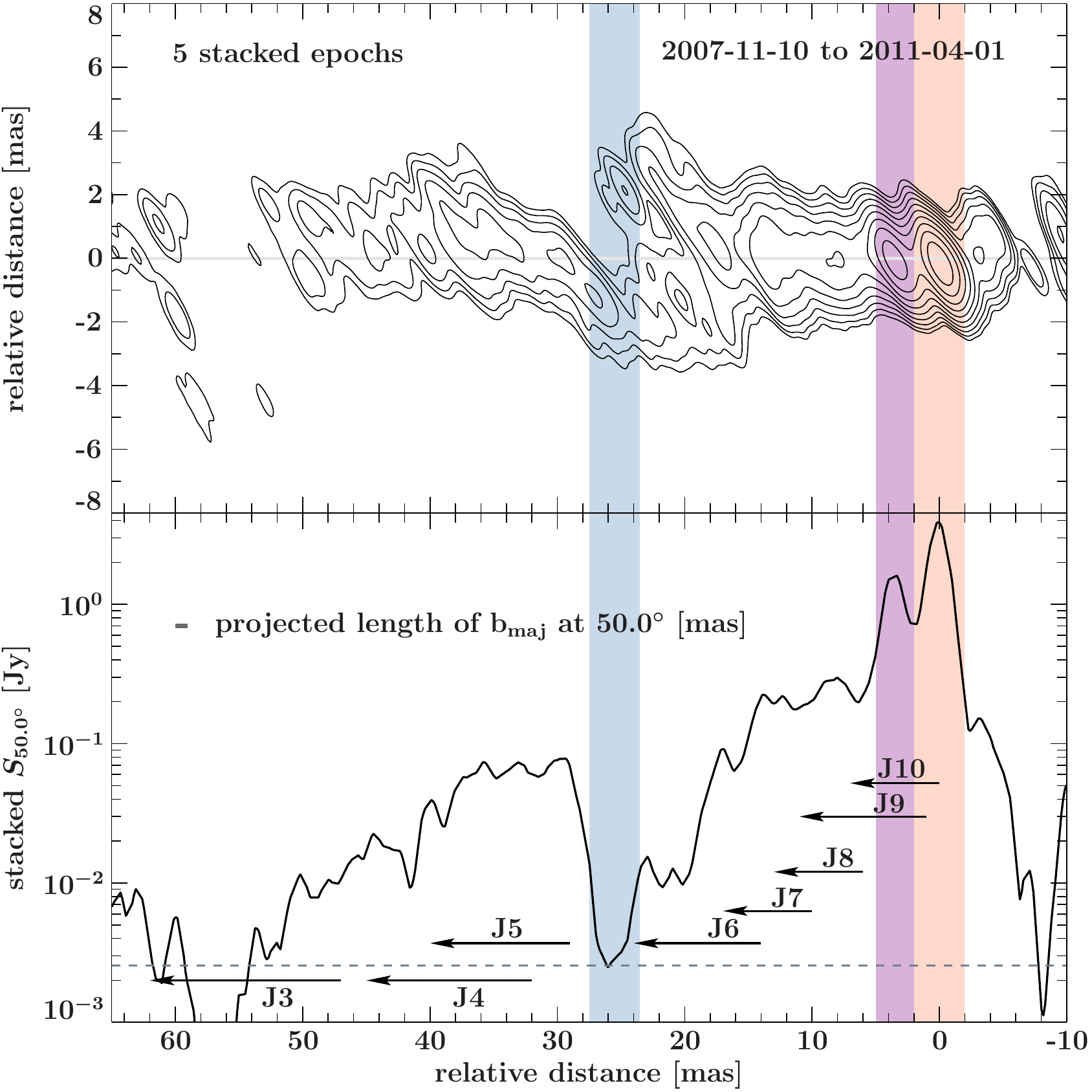}
\caption{\textit{Top:} Stacked \texttt{CLEAN} image. The five highest
resolution TANAMI images (epoch 2007.86, 2008.9, 2009.68, 2010.56,
2011.25) were restored with a common beam of size $(2.29 \times
0.58)$\,mas at $\mathrm{P.A.}=15.6^\circ$ and rotated by $40^\circ$.
\textit{Bottom:} Flux density profile along the jet axis (at
$\mathrm{P.A.}=50^\circ$) of stacked 8.4\,GHz \texttt{CLEAN} images.
The gray dashed line indicates the noise level of the stacked image.
The orange and purple shaded areas mark the core region and the stationary
component, respectively.  The blue shaded area at $25.5\pm2.0$\,mas
away from the phase center indicates the region of the jet where the
widening and decrease in surface brightness at 8\,GHz occurs. The
projection of the restoring beam width onto the jet axis is shown as a gray
line.  The black arrows indicate the traveled distances of the
identified components causing a smoothing of the profile due to the
jet flow with $\mu_\mathrm{mean}\sim
(2-3)\,\mathrm{mas}\,\mathrm{yr}^{-1}$ \citep[see
Sect.~\ref{susec:evolution} and compare to][]{Tingay2001b}. }
  \label{fig:fluxcuts}
\end{figure}

We can identify multiple individual components besides the core
(Fig.~\ref{fig:kinematic}), labeled J1 to J10 and $\mathrm{J_{stat}}$
(see Table~\ref{table:2}). One component is found to be stationary
($\mathrm{J_{stat}}$).  It shows remarkably constant brightness
temperature behavior. For the analysis of the time evolution of the
components, we excluded the whole region of the jet widening
(Sect.~\ref{susec:imaging}), since cross-identification of related
edge-components suffers from larger positional uncertainties and
larger offsets to the jet axis.  The position of the ``tuning fork''
like emission with lower surface brightness remains stationary, as
clearly seen in the stacked flux density profile
(Fig.~\ref{fig:fluxcuts}).

Seven of the moving components are detected in more than four
consecutive epochs. Two newly emerged components (J10 and J9) are
ejected into the jet during the TANAMI monitoring period. Component
J10 could only be detected in three epochs so far and further VLBI
observations are required to better constrain its trajectory (see
below). The two outermost components J2 and J1 only appear in two data
sets because of limited short-baseline $(u,v)$-coverage and
sensitivity in some epochs. The study of the jet kinematics of Cen~A
will be discussed based on the time evolution of the components J3 to
J10.

\begin{figure}
\includegraphics[width=\hsize]{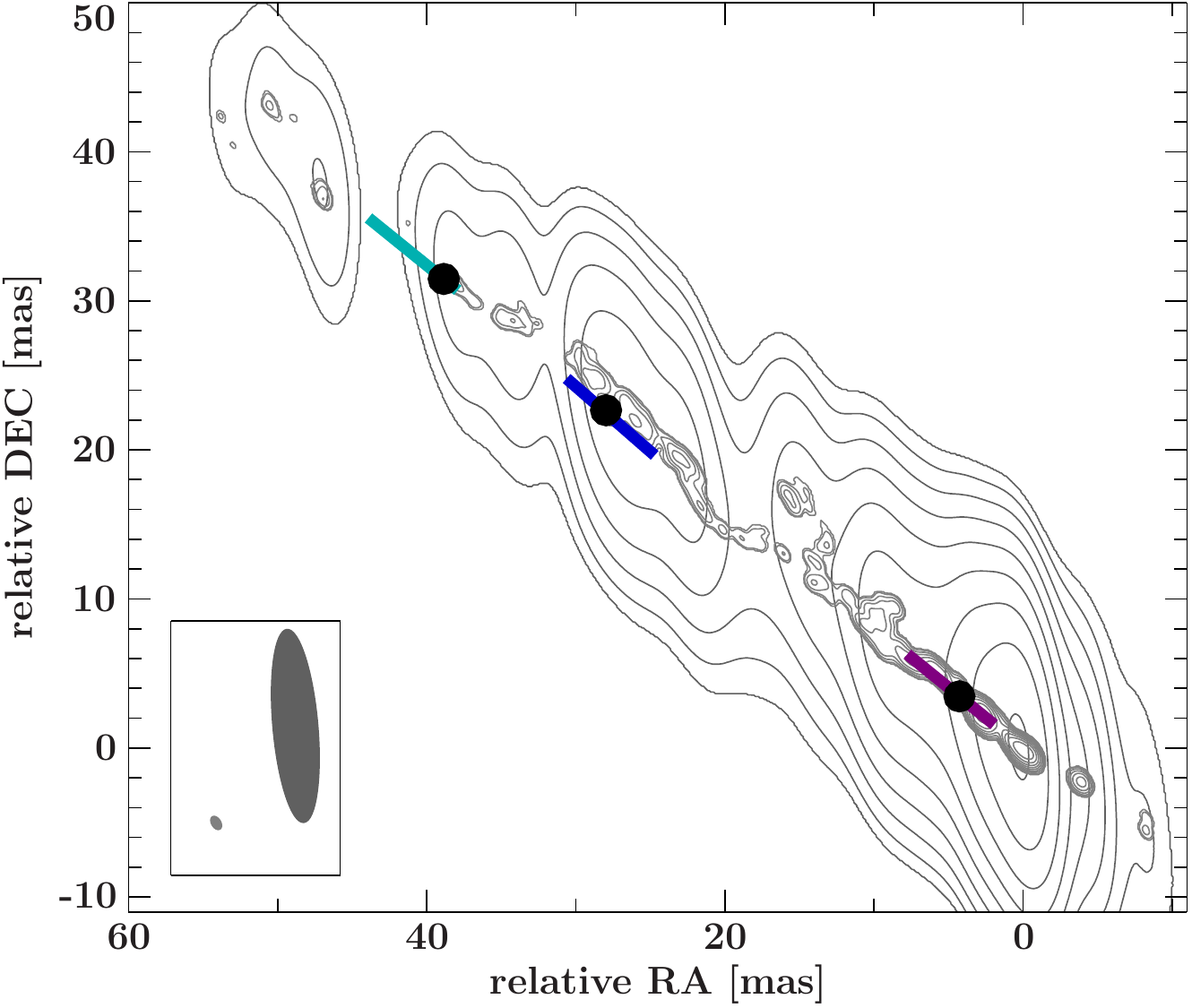}\hfill 
\caption{Tapered (dark gray) and original 2008 November TANAMI image
  (light gray). A taper of $0.1$ at $25\,M\lambda$ was chosen and the image
  was restored with the common beam of $(3\times 13)$\,mas at
  $5^\circ$ as applied in \citet{Tingay2001b}
   in order to compare the TANAMI
  observations with these previous VLBI measurements. In the lower
  left the size of the respective restoring beam is shown as a
  gray/black ellipse. The tapered TANAMI image is remarkably similar
  to the images by \citet{Tingay1998b,Tingay2001b}.
  The extrapolation of the motions of the components 
    monitored by \citet{Tingay2001b}, adopting the respective measured apparent speed
   for each of the three components C1, C2, and C3, results in hypothetical component
    positions for this observation. The black filled circles indicate
    these expected positions, the purple/blue/cyan lines represent the positional
    uncertainties of C3/C2/C3 derived from the velocity error. This comparison shows
    that TANAMI observations are consistent with previous results.}
  \label{fig:taper}
\end{figure}

\begin{figure}\centering
\includegraphics[bb= 5 5 420 1100, clip, width=0.7\hsize]{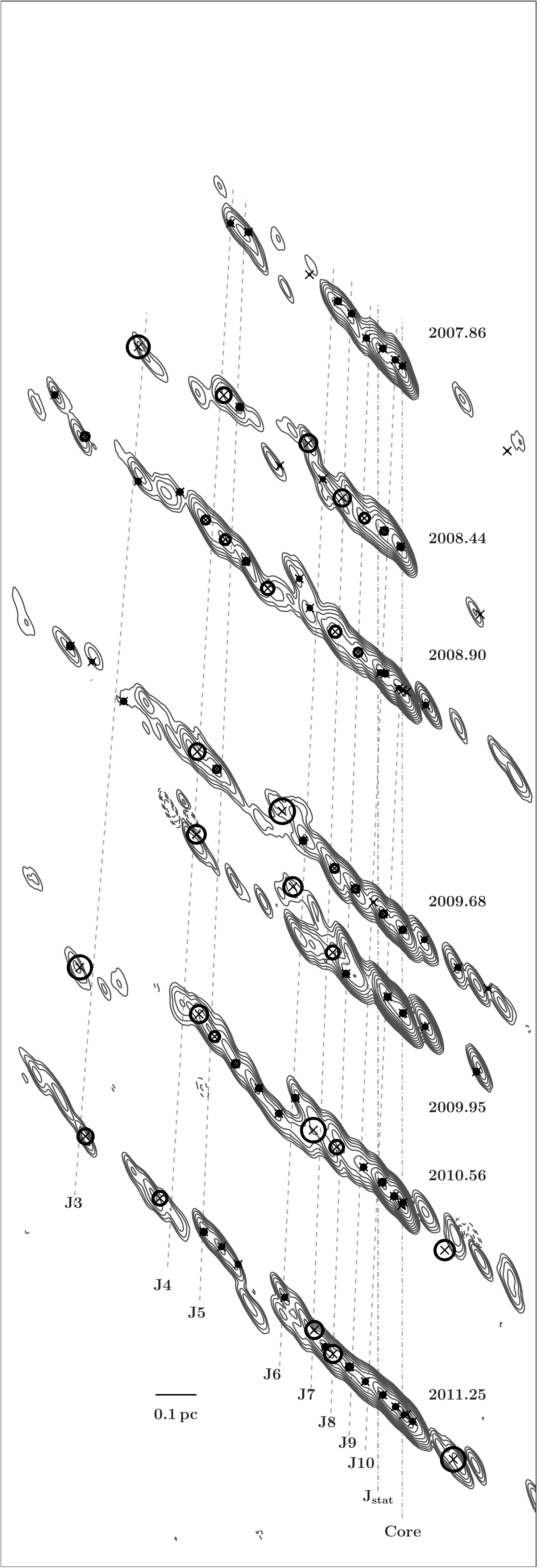}
\caption{Time evolution of Cen~A at 8.4\,GHz. Contour clean images,
restored with a common beam of $(3.33 \times 0.78)$\,mas at
P.A.=$26.3^\circ$. The contours indicate the flux density level,
scaled logarithmically, and increased by a factor of 2, with the lowest
level set to the $5\sigma$-noise-level. The positions and FWHMs of the
Gaussian \texttt{modelfit} components are overlaid as black circles (for model
parameters see Table~\ref{table:2}).
  } \label{fig:timeevolution}
\end{figure}

\begin{figure} 
\includegraphics[width=\hsize]{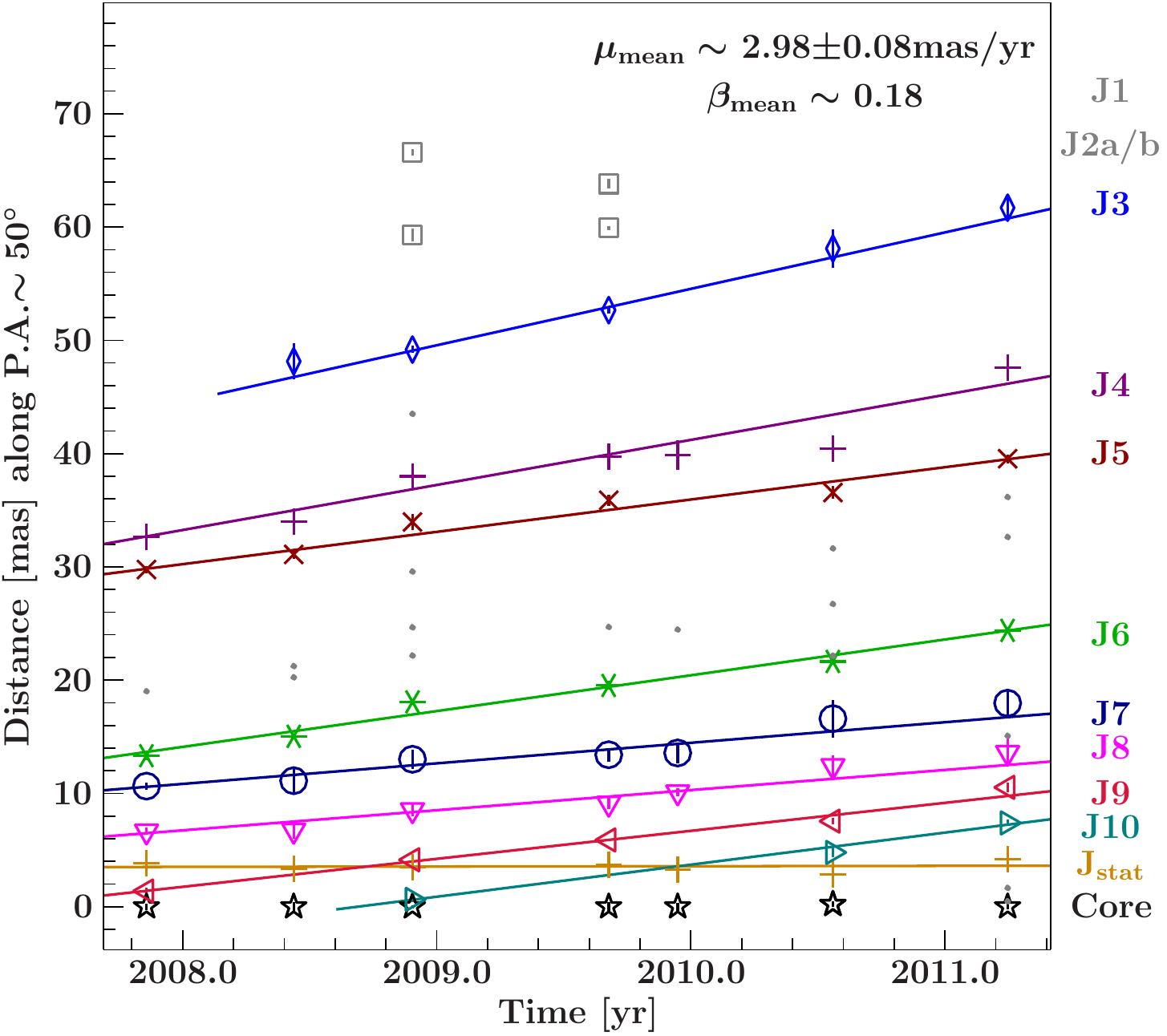} 
\caption{Component distance from the core component (set to zero) as a
function of time (see Table~\ref{table:2} for model
parameters). For the identified components (colored symbols), the
error bars show the systematic errors, which are defined as $0.5\times
b_\mathrm{maj}$ for resolved components. For unresolved components, the
resolution limit is used
\citep[][]{Kovalev2005}. A linear regression
fit to associated components is
shown. Components not included in the fit are shown in gray. We note, in
particular, the components in the region of the ``tuning fork'' are
excluded since the region is too complex. The mean apparent speed is
determined for components J3 to J10.}
\label{fig:kinematic}
\end{figure}

\begin{figure}
\includegraphics[width=\hsize]{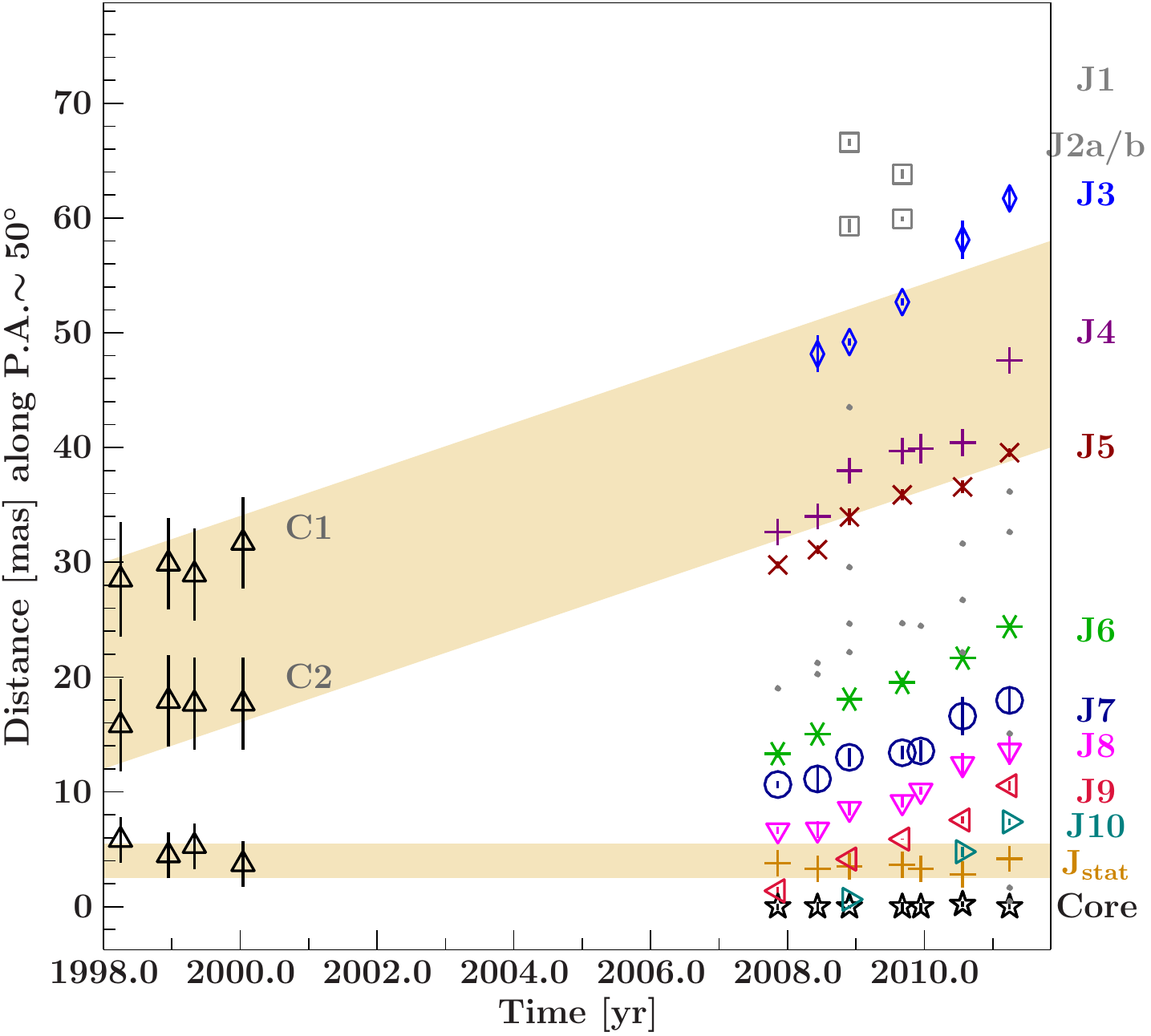}
\caption{8.4\,GHz kinematic results of Cen~A by \citet{Tingay2001b}
  (open triangles) and the 2007--2011 TANAMI data. The shaded region
  marks the expected velocity evolution using
  $\mu=2\,\mathrm{mas}\,\mathrm{yr}^{-1}$, which is comparable to the
  mean apparent speed given by the tapered TANAMI images (see text for
  details).}
  \label{fig:kinematic_tingay}
\end{figure}

Assuming ballistic motion for each component, we determine the
individual apparent speeds for the components J3 to J10 using linear
regression fits to the centroid position of associated Gaussians. This
is the most solid approach because testing for possible acceleration of
single components is not reliable until robust detection of components
in more than ten consecutive epochs, as discussed in
\citep{Lister2013}. We detect a clear velocity dispersion (see
Table~\ref{table:3}). The individual component proper motions range
from $\mu=1.78\pm0.19\,\mathrm{mas}\,\mathrm{yr}^{-1}$ for J8 to
$\mu=4.98\pm0.38\,\mathrm{mas}\,\mathrm{yr^{-1}}$ for J3.
\begin{table}
\caption{Apparent speeds of individual jet components}             
\label{table:3} 
{\centering
\begin{tabular}{llllll}
 \hline
 \hline
 ID & $\mu [\mathrm{mas/yr}]$   & $\beta_\mathrm{app}$    & $d_\mathrm{mean}
 [\mathrm{mas}]$\tablefootmark{a} &  $t_\mathrm{ejection}$\\
 \hline
 J3 & 4.98 $\pm$ 0.38 & 0.29 $\pm$ 0.02 & 53.96 &$\sim$1983\tablefootmark{b}\\
 J4 & 3.98 $\pm$ 0.23 & 0.23 $\pm$ 0.01 & 38.89 &$\sim$1989\tablefootmark{b}\\
 J5 & 2.85 $\pm$ 0.13 & 0.17 $\pm$ 0.01 & 34.47 &$\sim$1989\tablefootmark{b}\\
 J6 & 3.16 $\pm$ 0.10 & 0.19 $\pm$ 0.01 & 18.67 & 2002.0$\pm$1.0\\
 J7 & 1.82 $\pm$ 0.25 & 0.11 $\pm$ 0.01 & 13.76& 2002.0$\pm$2.0 \\
 J8 & 1.78 $\pm$ 0.19 & 0.10 $\pm$ 0.01 & 9.60 & 2005.0$\pm$1.0\\
 J9 & 2.47 $\pm$ 0.13 & 0.15 $\pm$ 0.01 & 5.90& 2007.5$\pm$0.5 \\
 J10 & 2.83 $\pm$ 0.14 & 0.17$\pm$  0.01 & 4.27 & 2008.5$\pm$0.5 \\
 \hline
\end{tabular}
}\\
\tablefoot{\tablefoottext{a}{Mean component distance from the
    core. Note that the time range over which single components have been
tracked is shorter for the two newly emerged components J9 and J10.} 
\tablefoottext{b}{Results from \citet{Tingay2001b}, see main text for details.}}
\end{table}
The mean apparent speed of these eight components is
$\mu_\mathrm{mean}=2.98\,\mathrm{mas}\,\mathrm{yr}^{-1}$, and
the median is
$\mu_\mathrm{median}=2.83\,\mathrm{mas}\,\mathrm{yr}^{-1}$.  This shows
a broader range of speeds for the different resolved components than
the values determined by \citet{Tingay2001b}.  The tapered data can
explain this discrepancy.  As discussed in Sect.~\ref{susec:taper},
the comparison of the tapered TANAMI images with these previous
observations and the extrapolation of the component position to the
time of the TANAMI observations (see Figs.~\ref{fig:taper} and
~\ref{fig:kinematic_tingay}) allows us to identify the component
complex J5 to J3 with components C2 and C1 of \cite{Tingay2001b},
respectively (see Sect.~\ref{sec:discussion} for further details). We
can therefore conclude that the larger scale structure, shown in the
tapered images, moves with a mean speed comparable to the results by
\cite{Tingay2001b}.  This comparison shows the influence of limited
angular resolution in Cen~A kinematic studies. Higher resolution
TANAMI observations allow us to detect small-scale structures that
seem to follow a pre-existing channel defined by the flow, but show
distinct apparent speeds.

Assuming constant component velocities and no intrinsic acceleration,
the back-extrapolation of the component tracks constrains the ejection
times for the innermost components (see Table~\ref{table:3}).  The
ejection times of J5 to J3 are obtained using the tapered analysis and
the cross-identification with C2 ($1989.2^{+0.9}_{-0.7}$) and C1
($1983.5^{+2.2}_{-3.2}$) from \citet{Tingay2001b}.

\begin{figure}
\resizebox{\hsize}{!}{\includegraphics{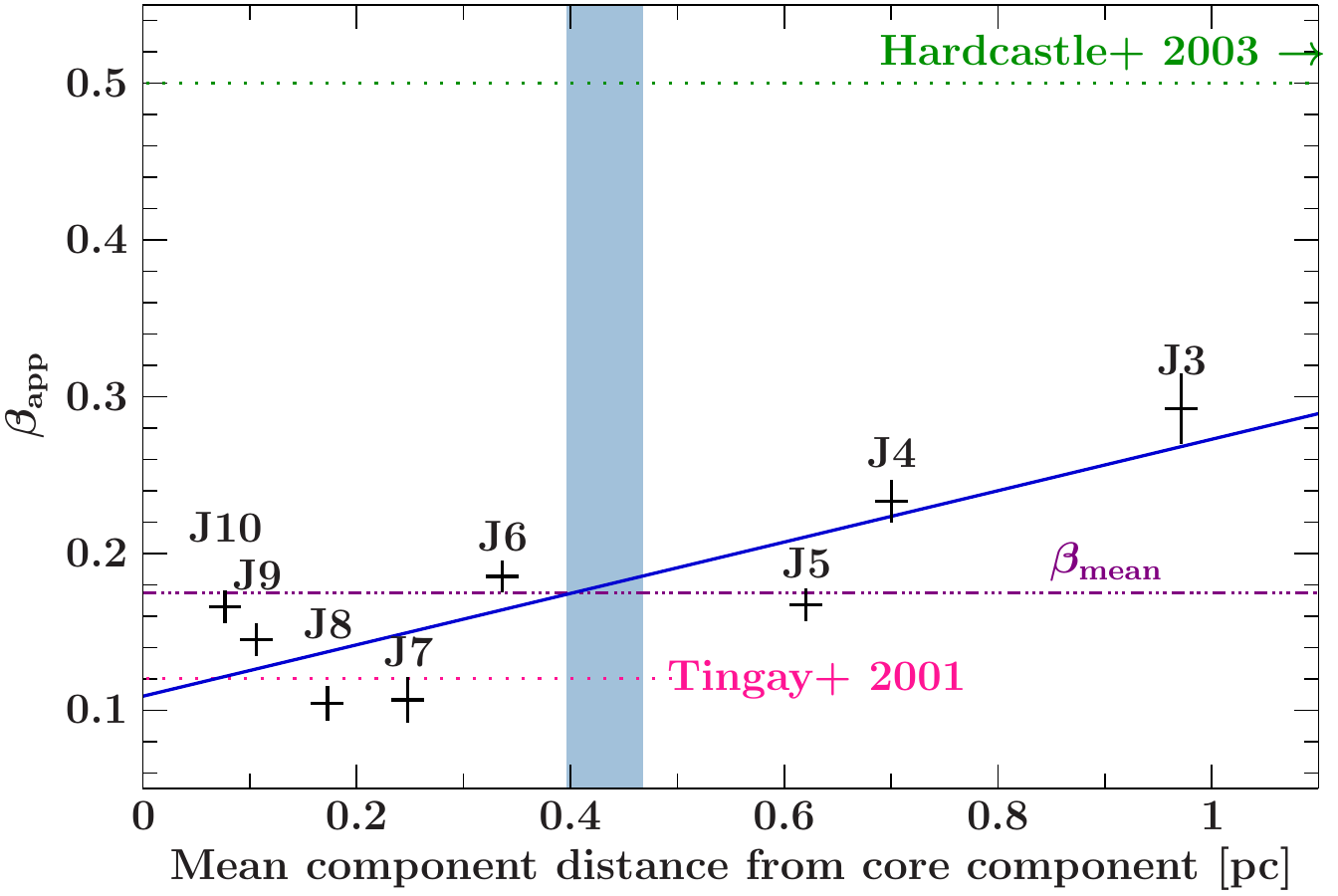}}
\caption{Evolution of individual component velocities as a function of
mean component distance from the core component, which can be
  parameterized by a
  linear fit of $\beta_\mathrm{app} = 0.16\,d + 0.11$ (blue line). The blue shaded area
marks the position and extent of the jet widening
(Sect.~\ref{sec:discussion}). The mean component speed and archival
measurements by \citet{Tingay2001b} and \citet{Hardcastle2003} are
indicated by straight lines. }
  \label{fig:vdist}
\end{figure}

Figure~\ref{fig:vdist} shows the $\beta_\mathrm{app}$-distribution for
each component as a function of mean distance from the core.  The
pc-scale jet of Cen~A shows a trend toward an increasing component velocity
farther downstream, which can be parameterized by a linear fit of
$\beta_\mathrm{app} = 0.16\,d + 0.11$, where $d$ is the mean component
distance in pc over the observed time range. Components J4 and J3 show
significantly higher speeds than components closer to the core,
suggesting that the jet undergoes acceleration farther out, as seen in
a statistically large sample of AGN jets by \citet{Lister2013}.

The best-fit apparent velocities of the two newly emerged components
J9 and J10 agree within their errors, but deviate from the mean jet
speed. Both components pass the stationary feature during the time of
TANAMI monitoring, a very complex region where misidentification is
possible. They are faster than the more robust components
J7 and J8. More VLBI observations will be able to test the higher
speeds of J9 and J10.

\subsection{Simultaneous ballistic fit}\label{susec:simfit}
The characterization of the jet flow with ballistic tracks is a good
approach to determine the motion of the individual
components. However, the motion tracks could in principle be more
complex.  In the following we introduce a method that fits all
data sets for all epochs simultaneously with the constraint that
component trajectories have to follow ballistic tracks. We couple the
positions of all components in the individual epochs using a single
linear model and then solve for all component velocities by performing
a simultaneous fit to all epochs. 
As a starting model, to identify the components we use the
models obtained for each single $(u,v)$-dataset. When assuming ballistic
motion, the time evolution of each identified \mbox{component $i$} (with
$i=1\ldots10$) can then be described by
\begin{equation}
\begin{aligned}
   x & = x_{0,i} + \cos(\phi_i)  v_i  (t - t_{0,i})\\
   y & = y_{0,i} + \sin(\phi_i)  v_i  (t - t_{0,i})
\end{aligned} 
\end{equation}
where $(x_0,y_0)$ is the the starting position, $\phi_i$ is the ejection
angle, $t_{0,i}$ the  ejection date, and $v_i$  the
velocity. 
Based on this input model
we performed a simultaneous fit to all datasets by fitting ballistic
motion to all identified components, resulting in a combined $\chi^2$.
As a consequence of this approach, if the motion is non-ballistic, the
fit will result in a high $\chi^2$-value, so that we can reject the
hypothesis of simple ballistic motion.

This approach results in a $\chi^2 = 74149$ with $39044$ degrees of
freedom ($\chi_\mathrm{red}^2 = 1.9$) for the combined, simultaneous
fit. The resulting component velocities agree with the
values obtained by the single fits using the ``classical'' approach
(see Sect.~\ref{susec:evolution}).  We conclude that the motion of the
identified components are described well by ballistic motion on the
basis of seven consecutive observations, though the best-fit has only
a moderate $\chi_\mathrm{red}^2$-value.  The downstream acceleration
is confirmed by this analysis, with $\sim
1-2\,\mathrm{mas\,yr^{-1}}$ faster speeds for J5 to J3 than for
the inner components.  We expect that with further monitoring epochs,
this analysis will yield a good test of a continuous acceleration
model along the jet.
 
\section{Discussion}\label{sec:discussion}

In the following sections we discuss the Cen~A jet kinematics. On top
of the overall jet flow, the series of highly resolved images of Cen~A
reveal particular bright and interestingly shaped structures, which are
discussed in detail.

\subsection{Overall jet structure and flow}\label{susec:flow}
Our observations of Cen~A reveal a complex jet flow with a wide range
of individual component velocities
(Sect.~\ref{susec:evolution}). Although a significant component motion
is measured, the observed features in the space-VLBI image of 2000
\citep{Horiuchi2006} show that the basic jet structure is stable over
years, forming a well-confined channel. The median apparent speed then
may be associated with the underlying flow that follows this
preformed jet structure. At the same time, owing to the interaction
between the jet and the ambient medium, individual jet features are
developed that can be slower or even stationary.
 
As shown by the tapered analysis, the TANAMI observations resolve the
substructure of prominent regions into components of a few light days
in size. \citet{Tingay2001b} reported  some dramatic flux
variability of components without major changes in apparent speed,
which already hinted at a finer substructure.

We measured a position-dependent acceleration (see
Fig.~\ref{fig:vdist}) of the outer components (beyond $\sim 0.3$\,pc),
which is very similar to the velocity distribution seen in the subparsec
region of \object{NGC\,1052} \citep[see Fig.~13 in][]{Lister2013}. If
one assumes that the acceleration continues downstream, it could be
the explanation for the apparent mismatch in jet velocity at a
distance of $\sim100$\,pc \citep[$\sim 0.5c$;][]{Hardcastle2003} and
the parsec-scale velocities measured by VLBI \citep[][and this
work]{Tingay2001b}.  However, when assuming the measured
acceleration continues downstream, the jet flow already reaches a
speed of $\beta_\mathrm{app}\sim 0.5$ at a distance of $\sim 2.5$\,pc.
Alternatively, jet bending within the inner few parsecs could also
explain the observed apparent acceleration.  \citet{Tingay1998b}
constrained the angle to the line of sight on parsec scales to
$50^\circ$--$80^\circ$, while the analysis at $\sim100$\,pc by
\citet{Hardcastle2003} resulted in $\sim 15^\circ$.  

\begin{figure}
\includegraphics[width=\hsize]{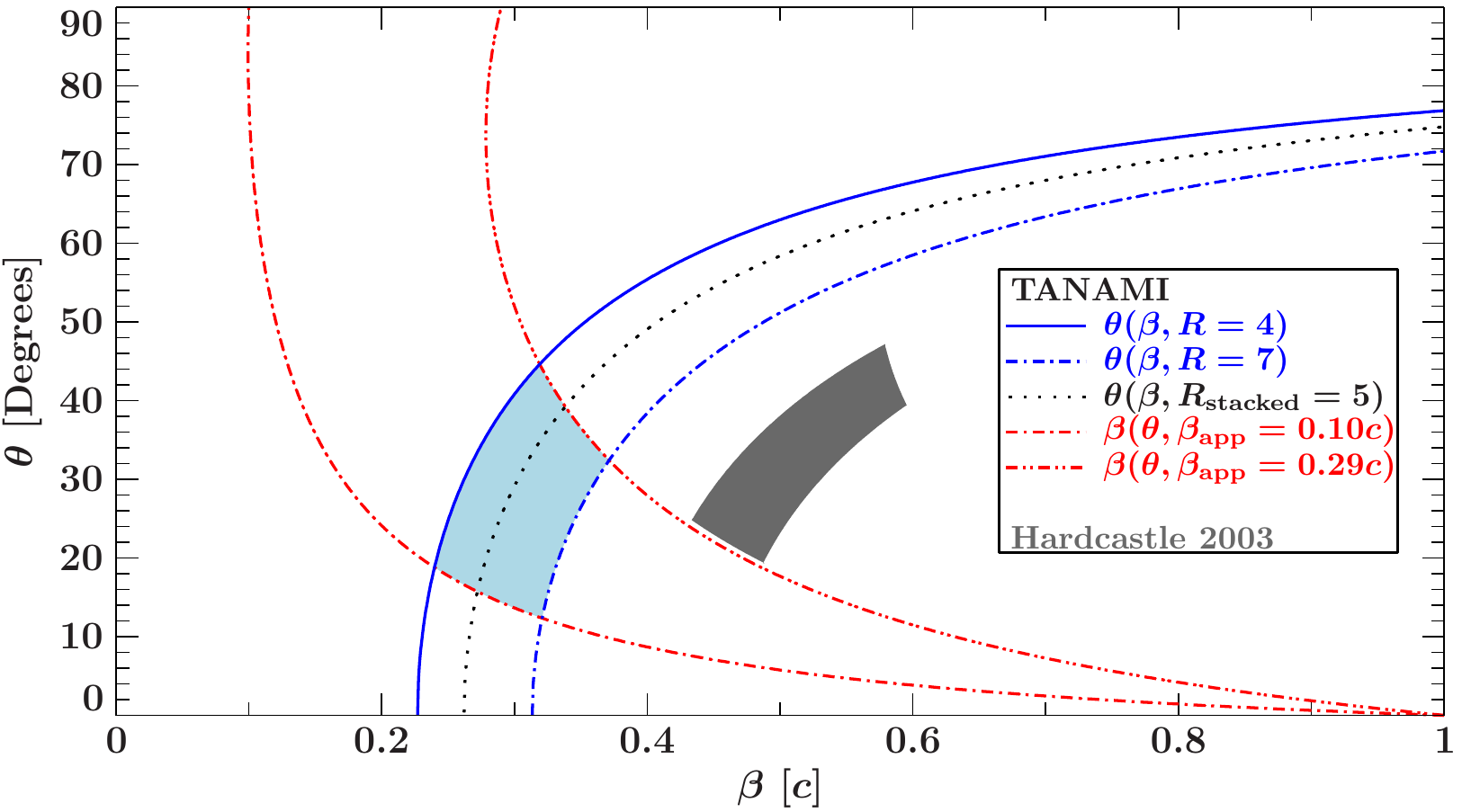}
\caption{Constraints on the values of the intrinsic jet speed $\beta$
  and the angle to the line of sight $\theta$ consistent with our
  measurements. Blue lines show the constraints on the $\theta$-range for
  the measured jet-to-counterjet ratio. The
  black-dotted curve gives the corresponding values for the stacked
  image. The red lines give the constraints on intrinsic jet speeds
  based on the measured apparent speeds of the slowest and fastest
  component (see Table~\ref{table:3}). The blue-shaded intersection area marks the
  region of permitted parameter space for both values. The gray-shaded
  region indicates the constraints according to measurements by \citet{Hardcastle2003}. \label{fig:betatheta} } 
\end{figure}

TANAMI observations yield surface brightness jet-to-counterjet ratios
of $R = 4$ and $R = 7$, using the two observations with the
best-sampled $(u,v)$-coverage of 2008 November 27 and 2011 April 1,
respectively, and excluding core emission between $-2\,\mathrm{mas}<
\mathrm{RA_{relative}}<2\,\mathrm{mas}$ to account for possible
absorption effects \citep[Paper~I;][]{Tingay2001a}. The stacked image
yields $R=5$.  Figure~\ref{fig:betatheta} shows the resulting
constraints on the angle to the line of sight $\theta$ and the intrinsic speed
$\beta$.  For an optically thin ($\alpha = -1$), smooth jet we use
\begin{equation}\centering
      R = \left(\frac{1+\beta \cos(\theta)}{1-\beta \cos(\theta)}\right)^{2-\alpha}\end{equation}
to constrain these parameters.  The measured brightness ratio, in
combination with the minimum and maximum measured proper motion, limits
the jet angle to the line of sight to \mbox{$\theta \sim 12^\circ - 45^\circ$}. Additionally,
we can constrain the intrinsic speed to \mbox{$\beta \sim 0.24 - 0.37$},
which is comparable to measurements for \object{NGC\,1052} \citep[$\beta=0.25$,][]{Vermeulen2003}.

Within the uncertainties, the upper limit of our derived angles is
consistent with the lower limits obtained by \citet{Jones1996} and
\citet{Tingay1998b}, while our lower limit overlaps with the values by
\citet{Hardcastle2003}.  These authors explained the discrepancy in
measured apparent speed between the parsec and the kilo-parsec scale
jet via a resolution effect and point out that \citet{Tingay1998b}
found hints of faster moving subcomponents.  Down to scales of about ten light days, we can exclude component speeds with
$\beta_\mathrm{app}\gtrsim0.3$, but find indications of downstream
acceleration.  We find that a larger angle to the line of sight cannot solely
explain the faster apparent speed measured by \citet{Hardcastle2003}, but
requires in addition an increase in $\beta$ from $\lesssim 0.3$ to
$\gtrsim 0.45$, since the angle is well
below the critical angle.  The higher apparent speeds on kilo-parsec
scales could thus consistently be explained with intrinsic
acceleration assuming a constant angle to the line of sight.

The spectral index distribution of the inner few light days of the jet
shows an optically thick core region with a steepening of the spectrum
downstream (Paper~I). It is striking that the region of the faster-moving components (J5 to J3) coincides with the optically thin region
of the jet.  A spine-sheath like jet-structure
\citep[e.g.,][]{Laing1999,Ghisellini2005,Tavecchio2008} could explain
this whole appearance, since a faster spine at small optical depth
would be revealed. For optically thick regions, only the outer --
slower -- layer could be probed. Further high-resolution spectral index
imaging is required, however, to confirm that coincidence.  A
stratified jet structure can also explain the different results on the
angle to the line of sight of Cen~A at parsec and kilo-parsec scales.
Observations of optically thin jet regions allow us to address the
faster and brighter spine, resulting in higher brightness ratios,
hence smaller estimated angles to the line of sight.  We conclude that
the different measurements of the angle to the line of sight
can be better reconciled by assuming a spine-sheath structure, with
possible intrinsic acceleration, rather than a large jet bending
within 100\,pc.

In principle, the observed downstream acceleration could also be
explained via an expansion of a hot jet in pressure equilibrium with
the ambient medium. If the ambient pressure has a steep density
gradient, the jet expands adiabatically and accelerates
\citep{Perucho2007}. In this scenario, standing, conical
(recollimation) shocks are expected to develop, though our
observations do not conclusively show such features (see
Sects.~\ref{susec:jstat} and~\ref{susec:fork}).

\subsection{The stationary component $\mathrm{J_{stat}}$.}\label{susec:jstat}
At a distance of $\sim 3.5$\,mas, the second brightest jet component
$\mathrm{J_{stat}}$ is found to be stationary, has an almost constant
brightness temperature, and is clearly resolved from the core
(Table~\ref{table:2}).  The spectral index of $\mathrm{J_{stat}}$ is
$\alpha \approx -1$, i.e., optically thin (Paper~I).
\citet{Tingay2001b} discussed the appearance of a stationary component
(C3) approximately at 4\,mas in mid-1993 with a weakly variable flux
density of about 1\,Jy (for a beam size of about
$(3\times13)$\,mas). This result suggests that $\mathrm{J_{stat}}$ is
a long-term stable feature.  Stationary components in extragalactic
jets are often observed and can be explained by locally beamed
emission or standing shocks
\citep[e.g.,][]{Piner2007a,Jorstad2001,Piner2007a,Lister2009c,Agudo2012}. The
alternative interpretation, a large jet bending at this position
\citep{Fujisawa2000}, had already been rejected by
\citet{Tingay2001b}, as the position angles of the
\texttt{modelfit}-components are all found to be similar. We confirm
this rejection based on our even higher resolution images.

We also note that while J9 and J10 pass through $\mathrm{J_{stat}}$,
neither of them shows significant evidence of interactions, e.g., a
flux change\footnote{However, due to the TANAMI observation
cadence, a short-time flux change might have been missed by our
observations.}.  A possible interpretation of $\mathrm{J_{stat}}$ is
that it represents a cross-shock in the jet flow as seen in
simulations of overpressured jets \citep{Mimica2009a}. In these
features a spectral inversion is expected, which is associated with
recollimation shocks, in which passing components are accelerated in
the following expansion. To confirm $\mathrm{J_{stat}}$ as such, a
recollimation shock will require further dual-frequency monitoring.
Given the measured steep spectrum, it is more likely that this feature
is the result of an internal, local pressure enhancement increasing
the density and emission, similar to a jet nozzle.

\subsection{The jet widening - the ``tuning fork''.}\label{susec:fork}

The ``tuning fork''-like structure at $\sim 25.5\pm 2.0$\,mas
(projected distance $\sim 0.4$\,pc) downstream is significantly detected in
multiple 8.4\,GHz TANAMI observations and is most prominently seen in the
highest dynamic range images.  The dip in the surface brightness
remains stationary, and the local jet morphology indicates a
circumfluent behavior (Fig.~\ref{fig:timeevolution}). According to
this result, we can conclude that we do not observe a moving
disturbance of the jet that is caused by a temporal disruption with
subsequent restructuring of the jet, but that this feature is likely a
standing discontinuity.

\citet{Tingay1998b} measured a strong increase in flux density of
component C1 when it reached the location of the ``tuning fork''. The
positional uncertainties of these previous VLBI measurements are not
comparable to our results but their Gaussian \texttt{modelfits} indicated an
increase in the diameter of C1 from $\sim$4\,mas to $\sim$\,14mas
within $\sim$3.5\,months.  \citet{Tingay2001b} reported on no strong
variations in the subsequent component C2; however, it did not reach
this location during their monitoring. We therefore conclude that the
jet widening might be a long-term stable feature such as
$\mathrm{J_{stat}}$.

At first glance the appearance of the ``tuning fork'' resembles a
recollimation shock creating a Mach disk that decelerates the jet
flow and separates it into two faster streams surrounding the central
Mach disk, as seen in simulations by \citet{Perucho2007}. In between
this ``tuning fork'' emission zone, i.e., behind the shock, one
expects optically thick emission and subsequent acceleration of the
jet flow.  While such a recollimation shock scenario could
describe the position-dependent acceleration in the succeeding
components J5 to J3, the overall appearance of the structure is
difficult to reconcile with the theoretical expectations, because when a Mach
disk is formed, one expects a strong jet expansion prior to the shock.
Additionally, after the Mach disk, the jet should undergo further
strong reconfinement shocks and be strongly decelerated. Observations
show, however, that it is still well collimated and not slowing down
up to a hundred parsecs \citep{Hardcastle2003,Goodger2010}.  Finally,
a recollimation shock would also create a bright standing feature
downstream of the Mach disk, which is not observed. Thus, we can
conclude the ``tuning fork'' is probably not caused by a recollimation
shock. 

A different explanation of the feature is a standing, local
disturbance that does not disrupt the entire jet, such as an
interaction with a massive object.  Figure~\ref{fig:fluxcuts} shows
that the jet surface brightness sharply increases behind the gap and
starts to decline again downstream.  This already suggests that the
jet flow is probably intermittent but not totally stopped.  In
particular, we consider the penetration of a cloud or a star, which is
highly likely in the inner parsec of the galaxy with estimated stellar
number densities of thousands per cubic-parsec close to the galactic
center \citep[e.g.,][]{Lauer1992,Bednarek1997}.  Following the
argument in \citet{Araudo2013}, the number of red giant stars
inside the volume covered by the inner-parsec of the Cen\,A jet can be
roughly estimated: using a jet diameter of $0.1$\,pc, an accretion
rate of $0.01\,L_\mathrm{edd}$, a stellar life time of $\sim1$\,Gyr,
and a black hole mass of $5.5\times 10^7\,\mathrm{M_\odot}$
\citep{Neumayer2010a}, one obtains on the order of $10^2$ low-mass
stars ($\lesssim 8\,\mathrm{M_\odot}$) inside the the inner-parsec of
the jet. It appears reasonable that a fraction of $\lesssim 1$\,\% of
those might be in the red-giant phase and interacting with the jet
during the TANAMI monitoring period. The exact numbers in these
estimations, however, strongly depend on the stellar population, the
initial mass function, and the star-forming history and are not well
constrained for the inner parsec \citep[see, e.g.,][for observations
and simulations of the recent star formation history and stellar
population in the halo of
Cen~A]{Soria1996,Harris1999,Rejkuba2001,Rejkuba2004,Rejkuba2011}.  For
further comparison, observations of Sgr~A$^\ast$ show stars orbiting
the SMBH at a distance of $\sim 15$\,light days \citep{Schoedel2002}.
Since Cen~A hosts a SMBH at its center that is about ten times more massive
than the one in the Galaxy, we can expect stars or dust clumps orbiting the
central engine down to distances of at least a few hundred light
days. The distance of the ``tuning fork'' matches this expected range,
with $\sim 25$\,mas corresponding to $\sim 600$\,light days.

Simulations by \citet{BoschRamon2012} describe the interaction of a
jet and a cloud or star orbiting the center of the AGN in the context of the
formation of high-energy flaring events
\citep{Bednarek1997,Barkov2010,Khangulyan2013}.
Such an interaction scenario between Cen~A's jet and stars of the
galaxy has already been discussed by \citet{Hardcastle2003},
\citet{Tingay2009}, and \citet{Goodger2010} to explain the radio and
X-ray knots in the hundred-parsec scale jet. The presence of a dusty
torus-like structure in the vicinity of the central black hole in
Cen~A is confirmed by measurements in the infrared and
X-rays. \citet{Rivers2011a} report an occultation event observed
during twelve years of X-ray monitoring with \textsl{RXTE} at a
distance commensurate with the molecular torus whose properties have
been adopted from IR measurements by \citet{Meisenheimer2007}. These
measurements confine the dust torus to a maximum extension of
$\sim0.6$\,pc ($\sim 700\,$light days) with an orientation axis
aligned with the jet axis, also matching the distance of the jet
widening from the jet core.

Interpreting the ``tuning fork'' as a stellar bow-shock structure due to
jet-star/cloud interaction, the VLBI observations constrain its size
to about $D_\mathrm{bow shock}\simeq 0.5\,\mathrm{mas} \simeq
0.01\,\mathrm{pc}$, with a jet diameter of $R_\mathrm{jet}\simeq
5\,\mathrm{mas} \simeq 0.1\,\mathrm{pc}$. The jet crossing time for
such an object orbiting the center of Cen~A with
$v \simeq 10^4\,\mathrm{km}\,\mathrm{s}^{-1}$ \citep[a typical
galactic rotation velocity, see][]{Araudo2010} is given by
\begin{equation}\label{eq:tj}\centering t_j \simeq 6.2\times 10^8
  \left(\frac{R_j}{3.1\times10^{17}\,\mathrm{cm}}\right)
  \left(\frac{v_c}{10^{9}\,\mathrm{cm}\,\mathrm{s}^{-1}}\right)^{-1}\,\mathrm{s}, 
\end{equation} 
which is approximately 20\,years. Assuming that the very middle of the
interaction process is in 2009/2010, where our VLBI images show that
the obstacle is clearly surrounded by the jet, we can thus set
1995-2005 as the decade in which the obstacle started entering into
the jet. 
This point in time is consistent with the end of monitoring by
\citet{Tingay2001b}. Their low-resolution images might indicate such a
feature.  The 3D simulations of a star or stellar wind
region entering a relativistic jet (Perucho \& Bosch Ramon, in prep.)
show that even when the obstacle starts to enter the jet, part of the
jet flow will already surround it on the outer side of the obstacle,
resulting in the circumfluent behavior observed in the VLBI images.

Taking into account that the size of the shocked region
($D_\mathrm{bow shock}\simeq 0.5\,\rm{mas}\,\simeq 0.01\,\rm{pc}$) can
be several times the size of the obstacle
\citep{BoschRamon2012}, the size of the obstacle can be estimated to
be about $0.1D_\mathrm{bow shock}$, i.e.,
$D_\mathrm{obstacle}\simeq 10^{-3}\mathrm{\,pc\sim 200\,AU}$.

With calculations by \citet[][Eq.~4]{Araudo2010}, we can determine a
lower limit on the cloud density\footnote{Estimations for a jet power
of $L_\mathrm{jet} \simeq 10^{44}\,\mathrm{erg s^{-1}}$
\citep{Abdo2010_cenacore}, a location of the cloud of $0.4$\,pc, a
cloud shocking time of 10\,years and a cloud radius $\simeq
10^{-3}$\,pc.} of $n_\mathrm{cloud} \geq 1.5\times
10^{10}\,\mathrm{cm^{-3}}$ (for more detailed calculations see
Appendix~\ref{sec:obstacle}), which is three magnitudes higher than the
values obtained for Cen~A by \citet{Rivers2011a}. Moreover, the
determined cloud velocity of the occultation event is about one
magnitude lower than the one we assumed. A hypothetical cloud with
similar parameters would be disrupted within one year, and the
bow-shock like structure would have disappeared if the penetration
time of a shock was smaller than the jet crossing time. Such an
event therefore cannot convincingly explain the persistence of the ``tuning
fork' structure over $\gtrsim 3.5$\,years. However, a
denser object like a red giant would meet the requirements, since this
scenario provides expected values and is thus favored by our
calculations.

The equilibrium point between the stellar wind and a relativistic jet
is determined by the wind and jet ram pressure. Following
\citet[][Eq.~2]{Komissarov1994},
\begin{multline}\label{eq:rs}
  R_\mathrm{s} = 2.3\times 10^{13}
  \left(\frac{\gamma_j}{5}\right)^{-1}\left(\frac{\dot{M}}{10^{-12}\,\mathrm{M}_\odot\,\mathrm{yr}^{-1}}\right)^{1/2}\\
  \left(\frac{v_\mathrm{w}}{10\,\mathrm{km}\,\mathrm{s}^{-1}}\right)^{1/2}
  \left(\frac{P_\mathrm{e}}{10^{-10}\,\mathrm{dyn}\,\mathrm{cm}^{-2}}\right)^{-1/2}
  \, \mathrm{cm},
\end{multline}
where $\gamma_\mathrm{j}$ is the jet flow Lorentz factor, $\dot{M}$ 
the typical mass loss rate of the stellar wind of an old late type
star (expected in elliptical galaxies), $v_\mathrm{w}$  the velocity
of the wind, and $P_\mathrm{e}$  the external pressure, considering
that the jet is close to pressure equilibrium with the ambient medium,
which is a reasonable assumption in the absence of strong
recollimation shocks. However, it is very difficult to give an
accurate number for this parameter, so it represents a source of error
in the determination of the properties of the wind.

With $R_\mathrm{s}=0.01\,{\rm pc}$ (bow shock diameter
$D_\mathrm{bs}$) and a jet flow velocity of $0.5\,c$
\citep{Hardcastle2003}, a typical stellar wind velocity of
$v_\mathrm{w}=100\,\mathrm{km}\,\mathrm{s}^{-1}$ and an ambient
pressure\footnote{We adopt here the values estimated at the core of
the radio galaxy 3C~31 from X-ray observations
\citep{Hardcastle2002}.}  of $P_e=10^{-10}\,\rm{dyn\,cm^{-2}}$, we
obtain an estimate for the mass loss rate of the star, $\dot{M}$:
\begin{multline} \dot{M} \simeq 10^{-8}
  \left(\frac{R_\mathrm{s}}{3.1\times10^{16}\,\rm{cm}}\right)^2
  \left(\frac{\gamma_\mathrm{j}}{1.15}\right)^2 \\
  \left(\frac{v_\mathrm{w}}{100\,\rm{km\,s^{-1}}}\right)^{-1}
  \left(\frac{P_\mathrm{e}}{10^{-10}\,\rm{dyn\,cm^{-2}}}\right)
  \rm{M_\odot\,yr^{-1}}.
\end{multline}
A red giant with a stellar wind of $v_\mathrm{wind}\simeq
100\,\mathrm{km\,s^{-1}}$ and a mass loss of $\dot{M}\simeq
10^{-8}\mathrm{M_\odot yr^{-1}}$ could create such an equilibrium
point at $R_\mathrm{s}=0.01$~pc if the jet flow has a velocity
$v_\mathrm{j}=0.5\,c$, preventing the approach of the jet flow to the
star during the whole crossing time of the star through the jet,
provided that the conditions in the wind and the jet do not change
dramatically.

Regarding the possibility of an obstacle, an increase in the
radio-emission of the region is expected when the interaction started.
From our estimations we conclude that the onset of the interaction
must have started between 1995 and 2005. High-energy emission is
expected to increase when the star enters the jet\footnote{The \textsl{CGRO}/OSSE and COMPTEL monitoring of Cen~A was between
1991 and 1995 \citep{Steinle1998}. At higher energies, a
$3\sigma$-detection by EGRET was reported \citep{Hartman1999}.}
\citep{BoschRamon2012,Barkov2010,Barkov2012}. In Sect.~\ref{susec:LC}
we discuss the flux density variability of Cen~A in various frequency
bands. In 2000/2001, an increase in emission in the radio and X-ray is
observed, but a mere coincidence cannot be excluded.
\citet{Araudo2010,Araudo2013} and \citet{Barkov2010} suggest that
$\gamma$-ray emission from jet-star interaction events in Cen~A could
be detectable.  Interestingly, after analyzing four
years of \textsl{Fermi}/LAT data, \citet{Sahakyan2013} report on a possible second
$\gamma$-ray spectral component in the core emission of Cen~A above
$\sim 4$\,GeV.  It is a further intriguing coincidence that the very
highest energy neutrino detected in three years of IceCube integration
between May 2010 and May 2013 \citep{bigbird} is positionally
coincident with Cen~A: at a positional uncertainty of $15.9^\circ$,
the offset is only $13.6^\circ$.  Cen~A is by far the brightest radio
source in the field and one of the strongest $\gamma$-ray sources. The
high-energy emission could be produced simultaneously by different
processes (including hadronic interactions, which could give rise to a
neutrino component), but it is difficult to disentangle them with our
current data.

Alternative explanations of the ``tuning fork'' involving, for example,
recollimation shocks or changes in the spine-sheath structure, are not
completely ruled out on the basis of the reported
observations. However, we conclude that the jet-star interaction
scenario can explain the observations well enough and is
particularly interesting from the perspective of testability.  If the
``tuning fork'' is caused by an obstacle moving with typical rotation
velocities, it should have gone through the edge of
the jet in 10 to 20 years. If it is a cloud, it could be completely
shocked and advected with the jet flow before that moment, depending
on its properties \citep{Araudo2010,BoschRamon2012}. If instead it is a star, we should expect it to cross the jet. In either
case, the situation is dynamic and, as such, temporary. If the
structure is generated by a stable recollimation shock or anything
else, it should last longer. With ongoing VLBI monitoring we will be
able to distinguish these scenarios.

\subsection{The outer jet flow -- components J5 to J3.}\label{susec:j5j3}

The time evolution of the outer components J5 to J3 reveals an
increase in apparent speed with distance from the core
(Fig.~\ref{fig:vdist}).  This coincides with the region where the jet
becomes optically thin (Paper~I).  As discussed in
Sect.~\ref{susec:flow}, the better $(u,v)$-coverage of the TANAMI
array allows us to resolve out bright emission regions of the jet,
picking up the substructure of the underlying jet flow. Assuming a
spine-sheath structure of the jet, the correlation of higher speeds
measured in optically thin regions could be explained consistently
without claiming an internal acceleration.  Furthermore, the tapered
analysis of the seven TANAMI observations of Cen~A allows us to
associate the prominent emission region composed of J3 and J4/J5, with
the previously defined and tracked components C1 and C2 by
\citet{Tingay2001b}.  This connection is the most natural overall
description of the long-term evolution of the jet.

However, as we see the position-dependent acceleration and can
model a clear substructure in this outer region of the jet, other
interpretation scenarios are plausible.
Regardless of the nature of the jet widening -- though there is most
likely an inter-relation -- the motion of the outer components can be
explained in the context of a forward shock triggering trailing
components.  \citet{Tingay2001b} discusses C2 and C3 as trailing
components of a forward shock (C1).  Following this interpretation, we
can directly compare the behavior of C1/C2 to J3/J4/J5.

Owing to a local perturbation in the jet, a major forward shock
component can be formed followed by trailing components
\citep{Agudo2001}. The simulations show an increase in velocity of the
trailing components, i.e., an acceleration as function of distance up
to the jet speed as a consequence of fluid expansion.  That $v_\mathrm{J5}\leq v_\mathrm{J4}\leq v_\mathrm{J3}$ fits this
picture.  A back extrapolation of the motion tracks of J5 and J4 gives
an intersection point at $\sim 25$\,mas from the core in early
2006. Trailing components of a forward shock can be easily
distinguished from ``normal'' jet ejections, since they are not
ejected from the core but created in the wake of the main perturbation
\citep{Agudo2001,Mimica2009a}.  If J3 is a forward shock, we can
constrain the jet proper motion to $v_\mathrm{J5}\leq v_\mathrm{jet}
\lesssim 4.98 \pm 0.38\,\mathrm{mas}\,\mathrm{yr}^{-1}$ (see
Table~\ref{table:3}), since the flow cannot be slower than the trailing
components.
 
As a result, J3 to J5 give a consistent picture of a forward shock and
trailing components.  In such a case the direct association of
J3 with C1 is difficult to reconcile, since the component speeds
differ significantly and internal acceleration would need to be
invoked.  Furthermore, the extrapolation of the tracks of C1 and C2
assuming a constant velocity of $\mu\sim 2$\,mas/yr
\citep{Tingay2001b} implies that the TANAMI observations should
capture at least the remnants of these components at the end of the
observable jet. These constraints support the more reasonable picture
of the detection and association of the emerging components, but
reject this proposed shock scenario.  The only way to resolve this
contradiction is the formation of a faster shock front (J3) on top of,
passing through, and exceeding in flux density, the underlying jet flow observed by
\citet[i.e., C1 and C2,][]{Tingay2001b}.  This can be easily tested by
future VLBI observations looking for further trailing components.

\subsection{Multiwavelength variability}\label{susec:LC}

Cen~A is a well-known high-energy emitter detected up to TeV energies
\citep{Aharonian2009} with a blazar-like broadband SED of the core
emission \citep{Abdo2010_cenacore}. Although intensely studied, the
origin and mechanisms behind the hard X-ray emission are still under
debate.  The soft X-ray observations by \citet{Kraft2002} and
\citet{Hardcastle2003} clearly show a contribution from the jet at
kilo-parsec scales. \citet{Evans2004a} suggest non-thermal X-ray
emission from the parsec-scale jet.  In the hard X-ray band, the
situation is still ambiguous. The observed emission is similar to
those in Seyfert galaxies, but it is not clear if the hard X-ray
origin is also the disk or corona, as in these systems
\citep[e.g.,][]{Markowitz2007}. Several results point to a jet-related
spectral component, linking (jet) activity to hardening of the X-ray
spectrum \citep{Tingay1998b,Fukazawa2011,Beckmann2011a}.

The TANAMI jet kinematics of Cen~A can help constrain the
emission origin of the hard X-rays further.  In principle, the correlation of
a jet ejection event with an active X-ray state would indicate a
similar emission origin. \citet{Tingay1998b} discusses the coincidence
of the ejection times of C1 and C2 with X-ray high states of Cen~A,
but this correlation study lacks a comparable, continuous monitoring
at these higher energies.

\begin{figure*}
\includegraphics[width=\hsize]{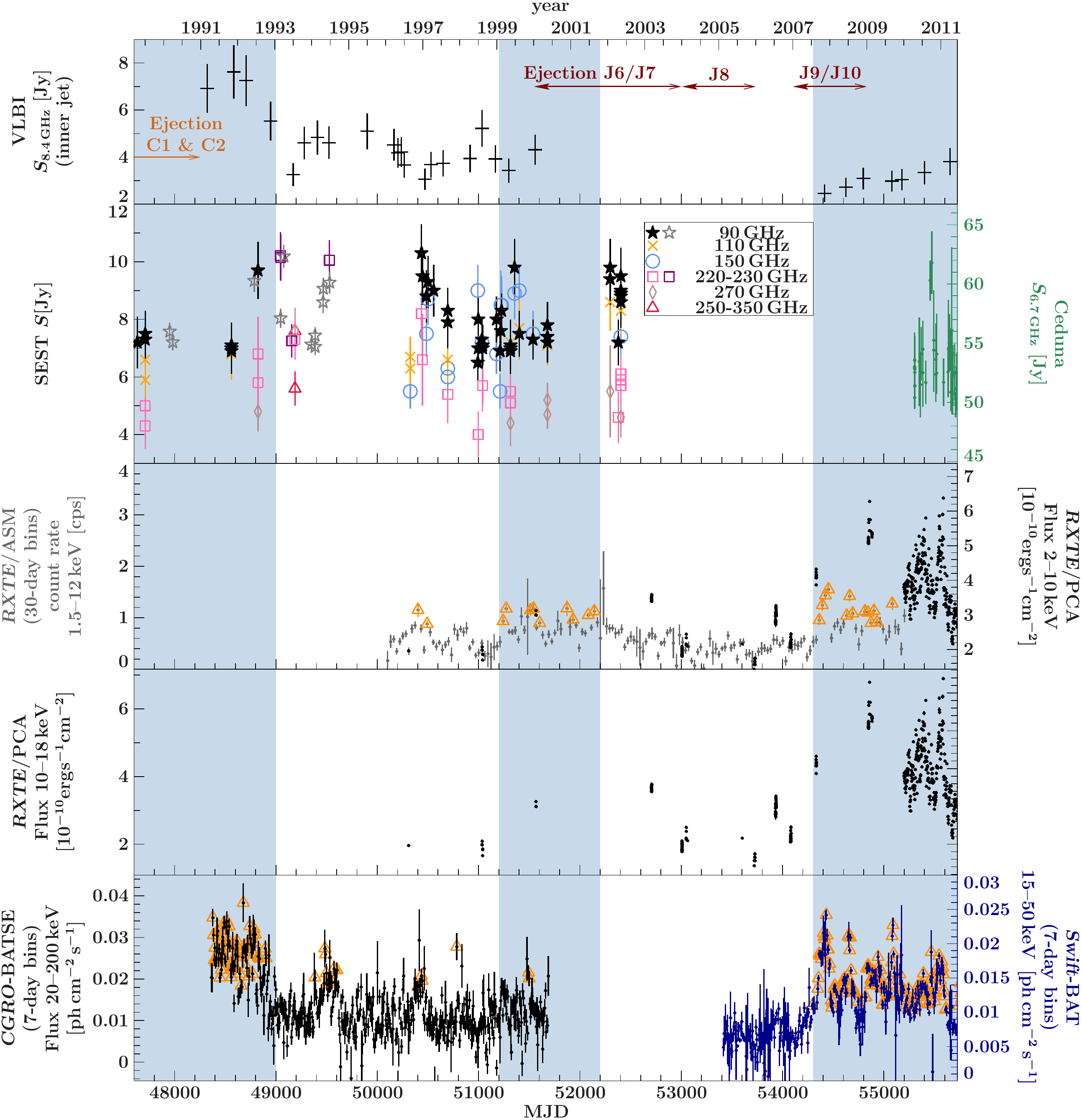}
   \caption{Radio and X-ray light curves of the Cen~A
core. \textit{From top to bottom}: Archival
\citep[][]{Tingay1998b,Tingay2001b} and TANAMI 8.4\,GHz flux density
of inner jet ($-2\,\mathrm{mas}\lesssim \mathrm{RA}_\mathrm{relative}\lesssim
15$\,mas) adopting an uncertainty of 15\%, ejection times including
uncertainties indicated by arrows (see Table~\ref{table:3}), SEST data
(90\,GHz -- 350\,GHz) from \citet{Israel2008} and
\citet{Tornikoski1996}, Ceduna monitoring at 6.7\,GHz,
\textsl{RXTE}/ASM \citep[Data taken after MJD 55200 are affected by an
instrumental decline and were therefore excluded; see,
e.g.,][]{Grinberg2013} and \textsl{RXTE}/PCA, \textsl{CGRO}/BATSE and
\textsl{Swift}/BAT data after background subtraction. The blue shaded
areas highlight the time periods of substantially prolonged higher
X-ray activity, defined by an X-ray flux of $3\sigma$ above the mean
(indicated by orange open triangles) in the continuous monitoring light curves by
\textsl{RXTE}/ASM, \textsl{Swift}/BAT, and \textsl{CGRO}/BATSE.}
  \label{fig:LC}
\end{figure*}

Figure~\ref{fig:LC} compares the archival X-ray light curves of continuous
monitoring by \textsl{Swift}/BAT, \textsl{RXTE}/ASM,
\textsl{RXTE}/PCA, and \textsl{CGRO}/BATSE with radio monitoring with SEST
at (90\,GHz to 350\,GHz) by \citet{Israel2008} and with the previous VLBI
\citep{Tingay2001b} and TANAMI monitoring results.  In the gamma rays, \textsl{Fermi}/LAT monitoring reveals Cen~A as a
persistent source without any flaring over more than two years
\citep{2fgl,1fhl}.  

The periods of long, persistently-high X-ray states are highlighted in
Fig.~\ref{fig:LC}. We also show the approximate ejection times of
individual radio knots. A partial overlap of the higher X-ray activity
and jet-component ejection is found, although it is not significant
enough to claim a common emission origin. The observation of more such
correlated events are required to confirm this result.  Furthermore,
as discussed in Sect.~\ref{susec:fork}, there is an indication of
higher X-ray activity in 1999 and 2000 with the possible onset of the
jet or obstacle interaction, though it is difficult to distinguish
between the contributions of different emission mechanisms causing
high energy activity.

It is, however, striking that the continuous increase in the radio flux density
of the inner jet (observed with TANAMI) follows the onset of activity
in hard X-rays from end of 2007 onwards, supporting the results by
\citet{Fukazawa2011} and following discussions by \citet{Evans2004a}. The
authors report on an additional power law component required to model
the hard X-ray spectrum at this high activity phase, possibly due to
increased jet emission. This is in contrast to spectral models
describing the hard X-ray spectrum during low X-ray activity
\citep{Markowitz2007}.  
Furthermore, note that the higher 8.4\,GHz
flux density in the inner jet follows the ejection of the
latest components J9 and J10. The radio flux rises simultaneously
with the detection of those components. As a result where the jet is becoming
optically thin and when the components are detected, the total flux
density increases, after the high X-ray activity.  A
similar correlation might be seen between the years 2001 and 2003
where the detection of J6 to J8 follows an X-ray high flux state.  The
SEST radio light curves from \citet{Israel2008} and
\citet{Tornikoski1996} do not allow us to identify a correlated radio
flare because of the sparse sampling and the large error bars.

We can conclude that there is a possible indication of a connection
between hard X-ray and jet activity in Cen~A.  The large
uncertainties in the ejection times mean that we cannot totally exclude the
possibility that Cen~A shows similar behavior to \object{3C\,120}
and \object{3C\,111} where jet ejections follow a dip in the X-ray
light curves \citep{Marscher2002,Marscher2006,Tombesi2012}, which can
be taken as proof of the accretion disk-jet interaction.  The reported X-ray dips are shorter than the phases of lower X-ray
emission seen in Cen~A. Our observations point instead to jet-related
hard X-ray activity than to such a direct proof of disk-jet
connection.  More ejection events and high-sensitivity X-ray spectra
are required to confirm this result.

\section{Summary and conclusions}\label{sec:conclusion}

We presented the jet evolution of Cen~A at highest angular resolution
ever obtained for this source, resolving and tracking up to ten
individual features over $\sim 3.5$\,years of less than 1\,mas in
size.  Connecting our results with previous VLBI studies, we obtain a
consistent picture of the central pc-scale jet.  The jet kinematics
can be explained by a spine-sheath structure with possible intrinsic
downstream acceleration. The similar appearance as a decade before
\citep{Horiuchi2006} in combination with a mean apparent speed, as
previously observed for larger structures \citep{Tingay2001b}, suggests
an underlying jet flow that is confined within a persistent
channel. Our observations clearly resolve this flow into individual
jet knots. Their apparent speeds suggest a downstream acceleration that
coincides with the optically thin jet.
 
This result connects to the findings of a faster jet at 100\,pc with a
spectral steepening in the outer jet regions, explainable by layers
with different particle acceleration conditions
\citep{Hardcastle2003,Worrall2008a}.  The jet-to-counterjet ratio
allows us to further constrain the angle to the line of sight on
parsec scales to $\sim 12^\circ - 45^\circ$, indicating a possible
intrinsic acceleration that further connects to the faster motion
detected on kilo-parsec scales.

We discussed the nature of stationary jet features, which persist
within the flow. The second brightest, stationary feature at a
distance of $\sim 3.5$\,mas from the core, is a long-term stable
component that has lasted in the jet for more than 15\,years. It is
most possibly due to a locally pressure maximum, similar to a jet
nozzle.  Farther downstream (at $\sim25.5$\,mas) a significant
decrease in the surface brightness is detected, accompanied by a widening of the
jet. This ``tuning fork'' structure can be
explained well by the jet hitting a star. This interaction causes a local
increase in the optical depth and forces the jet into a circumfluent
behavior, without an entire disruption. On the basis of the reported
observations, we cannot exclude the different explanation of this jet
feature as a recollimation shock. However, such interaction events are
expected for the denser populated central galactic region and have
already been proposed to explain jet knots at several 100\,pc
\citep{Hardcastle2003}.  It may be that such processes
contribute to the $\gamma$-ray emission detected from Cen~A
\citep{Abdo2010_cenacore,Sahakyan2013} or could even give rise to
neutrino emission.  Thanks to the expected short time range of such an
event, further VLBI monitoring will be able to test this scenario.

We observe a very dynamic jet with structural changes on timescales of
months to years. The long-term monitoring in the X-rays reveals
several relatively high flux states, prompting us to test for a
correlation to phases of stronger jet activity.  We can show that
the onset of the higher X-ray emission end of 2007 is followed by an
increase in the radio flux density of the inner jet contemporaneous
with detection of two new components. A similar coincidence is
indicated during active phases between 1999 and 2001 and was discussed
for the activity before 1992 by \citet{Tingay1998b}.  If such a
correlation of higher (hard) X-ray flux with higher jet activity is
confirmed by further VLBI monitoring, it will be a clear indication
of the emission origin of hard X-rays in Cen~A.  As discussed by
\citet{Fukazawa2011}, the high hard X-ray flux phase of Cen~A correlates with spectral hardening in the X-rays, possibly due to
increased jet emission. The time evolution of the innermost part of
Cen~A in radio to X-rays could then be explained by higher jet
activity causing an increase in high-energy emission.  This is
followed by a rise in radio brightness when these newly emerged
components become optically thin.  It is crucial to disentangle the
emission components in the X-rays in order to better constrain the
broadband spectral energy distribution of Cen~A.


\begin{acknowledgements}

The authors acknowledge the helpful discussions and suggestions by
A.~Lobanov and K.~Mannheim, and especially thank the anonymous referee
for valuable comments that improved the paper.

C.M. acknowledges the support by the ``Studienstiftung des Deutschen
Volkes''.  E.R. was partially supported by the Spanish MINECO projects
AYA2009-13036-C02-02 and AYA2012-38491-C02-01 and by the Generalitat
Valenciana project PROMETEO/2009/104, as well as by the COST MP0905
action ‘Black Holes in a Violent Universe’.  M.P. acknowledges
financial support by the Spanish ``Ministerio de Ciencia e
Innovaci\'on'' (MICINN) grants AYA2010-21322-C03-01,
AYA2010-21097-C03-01, and support by Universitat de Val\`encia and
Max-Planck-Institut f\"ur Radioastronomie.

The Australian Long Baseline Array is part of the Australia Telescope
National Facility, which is funded by the Commonwealth of Australia for
operation as a National Facility managed by CSIRO.  This research was
funded in part by NASA through {\em Fermi} Guest Investigator grants
NNH09ZDA001N, NNH10ZDA001N, and NNH12ZDA001N. This research was
supported by an appointment to the NASA Postdoctoral Program at the
Goddard Space Flight Center, administered by Oak Ridge Associated
Universities through a contract with NASA.  This research made use
of data from the NASA/IPAC Extragalactic Database (NED), operated by
the Jet Propulsion Laboratory, California Institute of Technology,
under contract with the National Aeronautics and Space Administration;
and the SIMBAD database (operated at the CDS, Strasbourg, France).

\end{acknowledgements}

\begin{appendix}
 \section{The ``tuning fork'' and a possible obstacle}\label{sec:obstacle}
In Sect.~\ref{susec:fork} we discuss possible explanations of the jet
widening (``tuning fork''), together with a decrease in surface brightness that appears
to be stationary over the TANAMI monitoring period. The jet flow is
likely to be disturbed by an obstacle. In the main text we conclude
that a scenario including a red giant with a significant stellar wind
is favored by our calculations and observations.  In the following, we
summarize the corresponding calculations when assuming the obstacle is a
cloud.

The shock produced in the cloud by the interaction with the jet should
propagate through it in a longer time than the penetration and
crossing time if we expect the cloud to cross the whole jet
diameter. Otherwise, the cloud would be disrupted by the shock when it
has completely crossed it, and the bow-shock structure would have
disappeared \citep{Araudo2010,BoschRamon2012}. Knowing that the
obstacle has been within the jet for at least ten years, hence a cloud
shocking time of $t_\mathrm{cs} > 3.1\times 10^8\,{\rm s}$, we can therefore
use Equation (4) in \citet{Araudo2010} to give a lower limit of the
density of the cloud:
      \begin{multline}\label{eq:tcs}
   t_\mathrm{cs} = \frac{2\,R_\mathrm{c}}{v_\mathrm{cs}} \\
\simeq 6.2\times 10^8 \left(\frac{R_\mathrm{c}}{3.1\times10^{15}\,\rm{cm}}\right) \left(\frac{n_\mathrm{c}}{10^{10}\,\rm{cm^{-3}}}\right)^{1/2} \\
   \left(\frac{z}{3.1\times10^{18}\,\rm{cm}}\right)\left(\frac{L_\mathrm{j}}{10^{44}\,\rm{erg\,s^{-1}}}\right)^{-1/2}  \, \rm{s},
   \end{multline}  
where we have taken the distance to the core $z=1\,{\rm pc} = 3.1\times 10^{18}$\,cm as a
reference. Taking the aforementioned lower limit on time into account
and taking as input data the obstacle radius $R_\mathrm{c}=10^{-3}$\,pc, the
jet power from the literature 
\citep[$L_\mathrm{j}\simeq 10^{44}\,{\rm erg\,s^{-1}}$,][]{Abdo2010_cenacore}, 
and the location of the ``tuning fork'' along the jet
$z\simeq 0.4\,{\rm pc}$, we can give a limit on the cloud density:
    \begin{multline}\label{eq:nc}
   n_\mathrm{c} > 1.5\times 10^{10} \left(\frac{t_\mathrm{cs}}{3.1\times10^{8}\,{\rm s}}\right)^{2}\left(\frac{R_\mathrm{c}}{3.1\times10^{15}\,{\rm cm}}\right)^{-2}\\
   \left(\frac{z}{1.2\times10^{18}\,{\rm cm}}\right)^{-2} \left(\frac{L_\mathrm{j}}{10^{44}\,{\rm erg\,s^{-1}}}\right)  \, {\rm cm^{-3}}.
   \end{multline}          
\citet{Rivers2011a} report an occultation of the central
core for 170\,days owing to the passage of a discrete clump of material
through the line of sight in the context of a clumpy torus model.  We
suggest that the eclipse is short because the angular size of the
eclipsed region must be small.  Interestingly, for the hypothetical
clump, which is located at a similar distance to the core to what is
reported for the location of the ``tuning fork'' \citep[$0.1\ldots0.3$\,pc,
adopting measurements by][]{Meisenheimer2007}, they obtain a central
density of $n_\mathrm{clump}=(1.8-3.0)\times 10^7\,{\rm cm^{-3}}$ and a
size of $R_\mathrm{clump}=(1.4-2.4)\times 10^{15}\,{\rm cm}$.  The size
of the clump would agree with our result, but the density we
obtain, as required for the cloud to survive the interaction for ten
years, is three orders of magnitude higher.  In addition, the cloud
velocity estimated by \citet{Rivers2011a} is one order of magnitude
lower than the one we used for our calculations
($v_\mathrm{clump}\simeq 1000\,{\rm km\, s^{-1}}$, Eq.~\ref{eq:tj}).
Such a decrease in the velocity would have two effects: 1.) the
crossing time would be increased by a factor of ten in Eq.~\ref{eq:tj},
i.e., $t_\mathrm{j}\simeq 6.2\times10^9\,{\rm s}$ ( approximately two
hundred years); and 2.) owing to this increase in the crossing time, the
lower limit in the cloud density obtained in Eq.~\ref{eq:nc} would be
increased by a factor of 100, bringing it to $n_\mathrm{c}>1.5\times 10^{12}\,{\rm
cm^{-3}}$.  Following the result of \citet{Rivers2011a}, we could thus
conclude that the cloud scenario should be ruled out for the ``tuning fork''
because: 1.) It is difficult to expect clumps with higher densities at
greater distances from the nucleus; and 2.) the shock crossing time of
the cloud, given by Eq.~\ref{eq:tcs}, would be reduced by a factor 20
with the numbers given in that paper for the cloud density,
thus giving $t_\mathrm{cs}\simeq3.4\times10^7\,{\rm s}$ (about one year). This
means that clumps of the size and density, such as those obtained by
\citet{Rivers2011a} for the case of a clumpy torus, would survive for
about one year before being disrupted and mixed with the jet if they
come to collide with it; i.e., they would be destroyed close to the
jet boundary, provided that their $t_\mathrm{j}$ is $\sim$200\,years.
Therefore, this scenario cannot explain the steady situation that is observed at
the ``tuning fork''.

 \begin{table}
 \caption{\texttt{Modelfit} parameters for individual jet components}             
 \label{table:2}
{\footnotesize
\begin{tabular}{llrrrr}
ID & $S$\tablefootmark{a} [Jy] & d\tablefootmark{b} [mas] &
$\theta$\tablefootmark{b} & $b_\mathrm{maj}$\tablefootmark{c}  & $\log
T_\mathrm{B}$\tablefootmark{d} \\ 
\hline \hline
\multicolumn{6}{l}{ \textbf{2007-11-10} ($\chi^2 = 5613.7$, d.o.f$=1954$)}\\
\hline
 & $0.01$ & $19.69$ & $-128.98$ & $<0.02$ & $>11.64$ \\ 
Core & $0.93$ & $0.00$ & -- & $0.58$ & $10.69$ \\ 
J9 & $0.54$ & $1.39$ & $48.68$ & $0.63$ & $10.39$ \\ 
$\mathrm{J_{stat}}$ & $0.61$ & $3.82$ & $48.31$ & $0.68$ & $10.37$ \\ 
J8 & $0.12$ & $6.66$ & $52.11$ & $0.62$ & $9.74$ \\ 
J7 & $0.11$ & $10.65$ & $44.13$ & $0.65$ & $9.66$ \\ 
J6 & $0.13$ & $13.33$ & $44.77$ & $0.72$ & $9.65$ \\ 
 & $0.02$ & $19.02$ & $45.41$ & $<0.02$ & $>11.64$ \\ 
J5 & $0.07$ & $29.77$ & $48.97$ & $0.65$ & $9.46$ \\ 
J4 & $0.06$ & $32.62$ & $50.25$ & $0.57$ & $9.47$ \\ 
\hline
\multicolumn{6}{l}{\textbf{2008-06-09} ($\chi^2 =3538.4$,
  d.o.f$=3828$)}\\
\hline
 & $0.04$ & $50.01$ & $-128.20$ & $2.11$ & $8.16$ \\ 
 & $0.03$ & $35.57$ & $-128.05$ & $0.49$ & $9.33$ \\ 
 & $0.03$ & $15.24$ & $-130.39$ & $<0.04$ & $>10.99$ \\ 
Core & $1.30$ & $0.00$ & -- & $0.79$ & $10.56$ \\ 
$\mathrm{J_{stat}}$ & $0.78$ & $3.33$ & $46.38$ & $0.98$ & $10.16$ \\ 
J8 & $0.39$ & $6.73$ & $51.66$ & $1.47$ & $9.50$ \\ 
J7 & $0.23$ & $11.12$ & $50.16$ & $2.41$ & $8.85$ \\ 
J6 & $0.06$ & $15.05$ & $48.99$ & $0.42$ & $9.80$ \\ 
 & $0.09$ & $20.25$ & $41.72$ & $2.47$ & $8.40$ \\ 
 & $0.03$ & $21.24$ & $56.13$ & $0.18$ & $10.27$ \\ 
J5 & $0.07$ & $31.10$ & $49.03$ & $0.82$ & $9.25$ \\ 
J4 & $0.10$ & $34.01$ & $49.41$ & $2.18$ & $8.59$ \\ 
J3 & $0.04$ & $48.14$ & $52.69$ & $3.20$ & $7.86$ \\ 
\hline
\multicolumn{6}{l}{ \textbf{2008-11-27} ($\chi^2 = 7878.1$,
  d.o.f$=7526$)}\\
\hline
 & $0.02$ & $86.13$ & $-128.75$ & $0.66$ & $8.94$ \\ 
 & $0.02$ & $79.90$ & $-127.60$ & $0.77$ & $8.81$ \\ 
 & $0.02$ & $52.37$ & $-129.00$ & $0.89$ & $8.57$ \\ 
 & $0.02$ & $35.57$ & $-124.72$ & $0.35$ & $9.45$ \\ 
 & $0.03$ & $21.71$ & $-126.62$ & $0.26$ & $9.83$ \\ 
 & $0.08$ & $4.02$ & $-123.25$ & $0.54$ & $9.70$ \\ 
 & $0.31$ & $0.64$ & $-118.58$ & $<0.02$ & $>11.82$ \\ 
Core & $0.73$ & $0.00$ &-- & $0.30$ & $11.17$ \\ 
J10 & $0.56$ & $0.64$ & $61.92$ & $0.39$ & $10.82$ \\ 
$\mathrm{J_{stat}}$ & $0.45$ & $3.50$ & $46.86$ & $0.73$ & $10.17$ \\ 
J9 & $0.24$ & $4.14$ & $52.20$ & $0.54$ & $10.16$ \\ 
J8 & $0.34$ & $8.53$ & $49.72$ & $1.18$ & $9.63$ \\ 
J7 & $0.28$ & $13.01$ & $49.15$ & $1.66$ & $9.25$ \\ 
J6 & $0.07$ & $18.08$ & $48.54$ & $0.51$ & $9.66$ \\ 
 & $0.04$ & $22.17$ & $42.88$ & $0.51$ & $9.47$ \\ 
 & $0.08$ & $24.66$ & $53.01$ & $1.82$ & $8.60$ \\ 
 & $0.11$ & $29.58$ & $50.56$ & $0.85$ & $9.41$ \\ 
J5 & $0.20$ & $33.96$ & $49.70$ & $1.44$ & $9.22$ \\ 
J4 & $0.12$ & $37.99$ & $49.20$ & $1.18$ & $9.17$ \\ 
 & $0.07$ & $43.51$ & $48.36$ & $0.67$ & $9.41$ \\ 
J3 & $0.04$ & $49.18$ & $51.73$ & $0.62$ & $9.21$ \\ 
J2 & $0.04$ & $59.30$ & $51.39$ & $1.15$ & $8.76$ \\ 
J1 & $0.03$ & $66.59$ & $49.68$ & $0.60$ & $9.10$ \\ 
\hline\hline
\end{tabular}
}
\tablefoot{All \texttt{modelfit} values are given without uncertainties, since
  they are dominated by systematics (see Sect.~\ref{susec:evolution}).
      \tablefoottext{a}{Integrated flux density of model component.}
     \tablefoottext{b}{Distance and position angle of the model component from the designated
      phase center.}
     \tablefoottext{c}{Major axis extent (FWHM) of the major axis.}
      \tablefoottext{d}{Logarithm of the brightness temperature of model component.}
}
\end{table}

\begin{table}            
{\footnotesize
\begin{tabular}{llrrrr}
ID & $S$ [Jy] & d [mas] & $\theta$ & $b_\mathrm{maj}$ & $\log T_\mathrm{B}$ \\ 
\hline \hline
\multicolumn{6}{l}{\textbf{2009-09-05} ($\chi^2 = 5250.2$, d.o.f$=7824$)}\\
\hline
 & $0.06$ & $44.71$ & $-130.13$ & $3.85$ & $7.83$ \\ 
 & $0.06$ & $32.40$ & $-125.85$ & $3.37$ & $7.93$ \\ 
 & $0.03$ & $20.19$ & $-123.59$ & $0.14$ & $10.47$ \\ 
 & $0.05$ & $15.13$ & $-124.53$ & $0.23$ & $10.23$ \\ 
 & $0.06$ & $9.72$ & $-124.34$ & $0.56$ & $9.56$ \\ 
 & $0.17$ & $3.47$ & $-114.55$ & $0.53$ & $10.03$ \\ 
Core & $1.62$ & $0.00$ & -- & $0.78$ & $10.66$ \\ 
 $\mathrm{J_{stat}}$ & $0.68$ & $3.67$ & $50.77$ & $0.95$ & $10.11$ \\ 
J9 & $0.04$ & $5.89$ & $46.78$ & $<0.03$ & $>11.98$ \\ 
J8 & $0.37$ & $9.12$ & $49.06$ & $1.03$ & $9.79$ \\ 
J7 & $0.24$ & $13.41$ & $47.98$ & $1.17$ & $9.48$ \\ 
J6 & $0.08$ & $19.53$ & $48.14$ & $0.79$ & $9.35$ \\ 
 & $0.10$ & $24.70$ & $45.47$ & $3.60$ & $8.14$ \\ 
J5 & $0.12$ & $35.88$ & $49.16$ & $0.98$ & $9.36$ \\ 
J4 & $0.15$ & $39.72$ & $49.05$ & $2.30$ & $8.71$ \\ 
J3 & $0.04$ & $52.66$ & $50.69$ & $0.64$ & $9.18$ \\ 
J2a & $0.02$ & $59.91$ & $49.17$ & $0.38$ & $9.30$ \\ 
J2b & $0.05$ & $63.81$ & $49.50$ & $0.82$ & $9.07$ \\ 
 & $0.03$ & $94.26$ & $51.31$ & $1.46$ & $8.43$ \\ 
\hline
\multicolumn{6}{l}{\textbf{2009-12-13} ($\chi^2 = 2232.0$,  d.o.f$=2350$)}\\
\hline
 & $0.05$ & $45.12$ & $-128.14$ & $3.24$ & $7.93$ \\ 
 & $0.09$ & $13.72$ & $-128.52$ & $0.63$ & $9.57$ \\ 
 & $0.35$ & $3.81$ & $-121.34$ & $0.52$ & $10.36$ \\ 
Core & $1.76$ & $0.00$ & -- & $0.72$ & $10.77$ \\ 
$\mathrm{J_{stat}}$ & $0.84$ & $3.27$ & $43.42$ & $0.72$ & $10.45$ \\ 
J8 & $0.31$ & $10.11$ & $55.31$ & $0.74$ & $9.99$ \\ 
J7 & $0.33$ & $13.56$ & $49.15$ & $1.89$ & $9.21$ \\ 
 & $0.12$ & $24.47$ & $41.03$ & $2.67$ & $8.48$ \\ 
J4 & $0.07$ & $39.89$ & $49.11$ & $2.59$ & $8.23$ \\ 
\hline\hline
\end{tabular}
}
\tablefoot{Table~\ref{table:2} continued.}
\end{table}

\begin{table}            
{\footnotesize
\begin{tabular}{llrrrr}
ID & $S$ [Jy] & d [mas] & $\theta$ & $b_\mathrm{maj}$ & $\log T_\mathrm{B}$ \\ 
\hline \hline
\multicolumn{6}{l}{\textbf{2010-07-24} ($\chi^2 = 5250.3$, d.o.f$=8642$)}\\
\hline
 & $0.02$ & $68.35$ & $-124.16$ & $4.42$ & $7.25$ \\ 
 & $0.07$ & $20.74$ & $-123.58$ & $4.32$ & $7.80$ \\ 
 & $0.06$ & $9.55$ & $-136.56$ & $3.08$ & $8.07$ \\ 
 & $0.06$ & $1.21$ & $-113.38$ & $<0.02$ & $>12.2$ \\ 
Core\tablefootmark{*} & $0.78$ & $0.22$ & $-44.02$ & $0.47$ & $10.79$ \\ 
Core\tablefootmark{*} & $0.75$ & $0.21$ & $135.98$ & $0.49$ & $10.74$ \\ 
$\mathrm{J_{stat}}$ & $0.69$ & $2.83$ & $46.61$ & $0.79$ & $10.29$ \\ 
J10 & $0.40$ & $4.79$ & $43.70$ & $0.89$ & $9.95$ \\ 
J9 & $0.19$ & $7.56$ & $47.80$ & $0.63$ & $9.93$ \\ 
J8 & $0.26$ & $12.43$ & $49.88$ & $1.92$ & $9.08$ \\ 
J7 & $0.34$ & $16.60$ & $50.64$ & $3.32$ & $8.73$ \\ 
J6 & $0.03$ & $21.66$ & $46.42$ & $0.63$ & $9.10$ \\ 
 & $0.09$ & $22.15$ & $54.23$ & $0.71$ & $9.48$ \\ 
 & $0.11$ & $26.72$ & $51.16$ & $0.70$ & $9.59$ \\ 
 & $0.08$ & $31.63$ & $49.86$ & $0.78$ & $9.35$ \\ 
J5 & $0.12$ & $36.57$ & $48.39$ & $1.12$ & $9.23$ \\ 
J4 & $0.12$ & $40.43$ & $47.11$ & $2.34$ & $8.57$ \\ 
J3 & $0.03$ & $58.10$ & $53.89$ & $3.41$ & $7.63$ \\  
\hline\hline
\multicolumn{6}{l}{\textbf{2011-04-01} ($\chi^2 = 23905.7$,  d.o.f$=6846$)}\\
\hline
 & $0.07$ & $111.32$ & $-126.36$ & $3.55$ & $7.96$ \\ 
 & $0.20$ & $55.99$ & $-123.57$ & $9.71$ & $7.56$ \\ 
 & $0.05$ & $39.70$ & $-124.57$ & $0.56$ & $9.48$ \\ 
 & $0.08$ & $36.29$ & $-122.40$ & $2.89$ & $8.20$ \\ 
 & $0.06$ & $23.94$ & $-124.06$ & $0.40$ & $9.83$ \\ 
 & $0.36$ & $9.75$ & $-131.91$ & $3.45$ & $8.73$ \\ 
 & $0.46$ & $1.65$ & $-127.32$ & $0.57$ & $10.40$ \\ 
 & $0.23$ & $0.48$ & $-81.91$ & $0.04$ & $12.41$ \\ 
Core & $0.85$ & $0.00$ & -- & $0.47$ & $10.84$ \\ 
 & $0.84$ & $1.66$ & $44.80$ & $0.62$ & $10.58$ \\ 
$\mathrm{J_{stat}}$ & $0.82$ & $4.19$ & $46.56$ & $0.65$ & $10.53$ \\ 
J10 & $0.31$ & $7.39$ & $49.16$ & $0.58$ & $10.21$ \\ 
J9 & $0.21$ & $10.54$ & $48.63$ & $0.88$ & $9.66$ \\ 
J8 & $0.23$ & $13.60$ & $49.24$ & $2.50$ & $8.82$ \\ 
 & $0.20$ & $15.09$ & $49.13$ & $0.74$ & $9.80$ \\ 
J7 & $0.21$ & $17.97$ & $46.48$ & $2.50$ & $8.76$ \\ 
J6 & $0.10$ & $24.39$ & $45.40$ & $0.58$ & $9.72$ \\ 
 & $0.06$ & $32.64$ & $47.70$ & $0.57$ & $9.53$ \\ 
 & $0.12$ & $36.16$ & $47.26$ & $0.57$ & $9.80$ \\ 
J5 & $0.14$ & $39.54$ & $47.53$ & $0.75$ & $9.64$ \\ 
J4 & $0.13$ & $47.58$ & $48.38$ & $2.02$ & $8.74$ \\ 
J3 & $0.15$ & $61.70$ & $48.80$ & $2.21$ & $8.73$ \\ 
 & $0.15$ & $105.84$ & $47.61$ & $3.60$ & $8.31$ \\ 
\hline\hline
\end{tabular}
}
\tablefoot{Table~\ref{table:2} continued.
\tablefoottext{*}{Due to the complex structure of the core
    region, a stable model requires two close components which are
    strongly correlated. Their flux weighted central point was taken as a
    reference position.}
}
\end{table}

\end{appendix}

\bibliographystyle{jwaabib}
\bibliography{mnemonic,aaabbrv,cena_kinematic}

\end{document}